\documentclass[prb,aps,twocolumn,showpacs]{revtex4-1}

\usepackage{graphicx,color}
\usepackage{amsthm}
\usepackage{amsfonts}
\usepackage{algorithmic}
\usepackage{enumerate}
\usepackage{latexsym}
\usepackage{amsmath}
\usepackage{amssymb}
\usepackage{bm}
\usepackage[pdftex,plainpages=false,colorlinks=true,linkcolor=blue, citecolor=blue,
urlcolor=blue]{hyperref}

\emergencystretch=\maxdimen
\begin{document}

%\title{{\small Quasiparticle scattering interference in the underdoped pseudogap phase of cuprate superconductors}}

\title{Unusual electronic ordering in the pseudogap phase of underdoped cuprate
superconductors}

\author{Xiang Li$^{1,2}$}
\author{Minghuan Zeng$^{3}$}
\thanks{E-mail: mhzeng@cqu.edu.cn}
\author{Yu Lan$^{4}$}
\author{Huaiming Guo$^{5}$}
\author{Shiping Feng$^{2,1}$}
\thanks{E-mail: spfeng@bnu.edu.cn}

\affiliation{$^{1}$School of Physics and Astronomy, Beijing Normal University,
Beijing 100875, China}

\affiliation{$^{2}$Department of Physics, Faculty of Arts and Sciences, Beijing
Normal University, Zhuhai 519087, China}

\affiliation{$^{3}$College of Physics, Chongqing University, Chongqing 401331, China}

\affiliation{$^{4}$College of Physics and Electronic Engineering, Hengyang Normal
University, Hengyang 421002, China}

\affiliation{$^{5}$School of Physics, Beihang University, Beijing 100191, China}

%\date{}

\begin{abstract}
The pseudogap phase of the underdoped cuprate superconductors harbours diverse
manifestations of different ordered electronic-states, and then these ordered
electronic-states coexist or compete with superconductivity. Here starting from the
microscopic electron propagator, the nature of the ordered electronic-states in the
pseudogap phase is investigated within the $T$-matrix approach. This $T$-matrix is
derived in terms of the inverse of matrix for various kinds of a single impurity,
and then is used to evaluate the local density of states (LDOS) by the involvement
of all the quasiparticle excitations and scattering processes. It is shown that a
number of the anomalous properties in the underdoped cuprate superconductors is
directly correlated to the opening of the normal-state pseudogap: (i) the structure
of the microscopic octet scattering model generated by the normal-state pseudogap
is essentially the same both in the superconducting (SC)-state and pseudogap phase,
which naturally leads to that the quasiparticle scattering interference octet
phenomenology observed in the SC-state exists in the pseudogap phase; (ii) however,
the spectral weight at around the antinodal region in the SC-state is gapped out
completely by both the SC gap and normal-state pseudogap, while it in the pseudogap
phase is suppressed partially by the normal-state pseudogap, this directly leads to
that the non-dispersive checkerboard charge ordering with a finite wave vector
${\bf Q}$ appears in the pseudogap phase only. The theory therefore also shows that
the electronic-states affected by the normal-state pseudogap exhibit the LDOS
modulation spectrum organization.
\end{abstract}

\pacs{74.62.Dh, 74.62.Yb, 74.25.Jb, 71.72.-h}

\maketitle

\section{Introduction}\label{Introduction}

In spite of over the past three decades of intense research, the cuprate
superconductors still hold many mysteries\cite{Keimer15,Varma20,Hussey23}. They are
complex systems that exhibit a variety of phases determined not only by temperature,
but also by the charge-carrier doping
\cite{Timusk99,Hufner08,Damascelli03,Campuzano04,Fink07,Hussey11,Vishik18,Drozdov18}.
This follows from a basic fact that the parent compounds of cuprate superconductors
are a Mott insulator\cite{Fujita12}, which originates from the strong electron
correlation\cite{Anderson87}, and then superconductivity with the exceptionally high
superconducting (SC) transition temperature $T_{\rm c}$ is obtained by doping charge
carriers\cite{Bednorz86,Wu87}, where the doping regimes have been classified into the
underdoped, optimally doped, and overdoped, respectively
\cite{Timusk99,Hufner08,Damascelli03,Campuzano04,Fink07,Hussey11,Vishik18,Drozdov18}.
However, this charge-carrier doping process nearly always introduces some measure
of disorder\cite{Hussey02,Balatsky06,Alloul09}, leading to that in principle, all
cuprate superconductors have naturally impurities, and then the electronic
inhomogeneity (then the inhomogeneous quasiparticle distribution) emerges naturally
\cite{Pan01,Fischer07,Yin21}. This remarkable evolution from a strongly correlated
Mott insulator with the localized electronic-state to a high-$T_{\rm c}$
superconductor with the delocalized electron pairing-state thus demonstrates
unambiguously that the anomalous properties of the electronic inhomogeneity in
cuprate superconductors are governed by both the strong electron correlation and
impurity scattering\cite{Zeng22,Zeng23}.

After intensive studies over three decades, the innumerable experimental observations
by using many probes have pointed out that the most of the exotic and unique features
of cuprate superconductors are to be found in the normal-state and where the deepest
mysteries lie
\cite{Timusk99,Hufner08,Damascelli03,Campuzano04,Fink07,Hussey11,Vishik18,Drozdov18}.
In particular, the normal-state in the underdoped regime is characterized by a
pseudogap\cite{Timusk99,Hufner08,Damascelli03,Campuzano04,Fink07,Hussey11} below the
pseudogap crossover temperature $T^{*}$ large compared to $T_{\rm c}$. In addition
to the high-$T_{\rm c}$ SC mechanism, this normal-state pseudogap is the most
noteworthy. The normal-state pseudogap depresses partially the electron density of
states on the electron Fermi surface (EFS), and then the anomalous properties
associated with the opening of the normal-state pseudogap are explained in a natural
way. This is also why in the underdoped regime, the phase above $T_{\rm c}$ but below
$T^{*}$ is labelled as {\it the pseudogap
phase}\cite{Timusk99,Hufner08,Damascelli03,Campuzano04,Fink07,Hussey11}. Although
the electronic inhomogeneity of cuprate superconductors in the pseudogap phase
is intrinsic in the nature, the quasiparticles scattering from impurities interfere
with one another, which leads to that the inhomogeneous quasiparticle distribution
is seeded by the impurity scattering
\cite{Pan01,Fischer07,Yin21,Zeng22,Zeng23,Pan00,Hoffman02,McElroy03,Kohsaka08,Hanaguri09,Vishik09}.
Experimentally, by virtue of systematic studies using scanning tunneling
microscopy/spectroscopy (STM/S) measurement techniques, a number of consequences
from the electronic inhomogeneity together with the associated fluctuation phenomena
in the pseudogap phase have been identified
as\cite{Vershinin04,Hanaguri04,McElroy05,Wise08,Lee09,Schmidt11,Fujita14}: (i) the
stronger quasiparticle scattering is detected at around the antinodal region than at
around the nodal region, which leads to that the constant energy contour for a finite
binding-energy is truncated to form the Fermi arcs centered at around the nodal
region with the largest spectral weight located at around the tips of the Fermi arcs,
which has been also confirmed by the angle-resolved photoemission spectroscopy (ARPES)
experiments\cite{Chatterjee06,McElroy06,Chatterjee07,He14,Restrepo23,Shah25}. The
characteristic features associated with this constant energy contour instability
are the emergence of the multiple nearly-degenerate electronic
orders\cite{Comin16,Comin14,Gh12,Neto14,Campi15,Hash15}, the dramatic change in the
line-shape of the energy distribution
curve\cite{Hash15,Dessau91,Campuzano99,Lu01,Sato02,Borisenko03,Wei08,Loret17,DMou17},
and the kink in the quasiparticle
dispersion\cite{Bogdanov00,Kaminski01,Johnson01,Sato03,Lee09-1,He13}. More specially,
these tips of the Fermi arcs connected by the scattering wave vectors ${\bf q}_{i}$
construct an {\it octet scattering model}, which can give a consistent description of
the regions of the highest joint density of
states\cite{Chatterjee06,McElroy06,Chatterjee07,He14,Restrepo23,Shah25};
(ii) the modulation of the local density of states (LDOS) in the SC-state appears for
the most part to be described by the quasiparticle scattering interference (QSI) caused
by the quasiparticle scattering from impurities, where the QSI peaks describe the
regions of the highest joint density of states, and are located at the well-defined
scattering wave vectors ${\bf q}_{i}$ (i=1,2,...,7) obeying the {\it octet scattering
model}; (iii) all these QSI signatures detectable in the
SC-state\cite{Pan00,Hoffman02,McElroy03,Kohsaka08,Hanaguri09,Vishik09}
survive virtually unchanged into the pseudogap phase. Concomitantly, all seven
scattering wave vectors ${\bf q}_{i}(\omega)$, which are dispersive in the SC-state,
remain dispersive into the pseudogap phase. In this case, the multiple
ordered electronic-states, which emerge in the pseudogap phase above $T_{\rm c}$,
persist into the SC-state below $T_{\rm c}$, and then they coexist and compete with the
SC-state\cite{Comin16,Comin14,Gh12,Neto14,Campi15,Hash15}; (iv) however, in addition to
the QSI octet phenomenology, the unusual checkerboard peaks in the LDOS modulation
at a finite scattering wave vector ${\bf Q}$ emerge in the pseudogap phase only. In
particular, this remarkable checkerboard charge ordering is observed to be non-dispersive,
and therefore is a static local charge ordering. These experimental observations therefore
provide the firm evidence for the LDOS modulation spectrum organization due to the opening
of the normal-state pseudogap.

Although the unusual features of the LDOS modulation in the pseudogap
phase\cite{Vershinin04,Hanaguri04,McElroy05,Wise08,Lee09,Schmidt11,Fujita14}
have been established experimentally, a complete understanding of these anomalous
properties is still unclear, where the crucial questions emerge in the field: (i) why the
characteristic of the QSI octet phenomenology observed in the SC-state also appears to
remain unchanged upon passing above $T_{\rm c}$ in the pseudogap phase? (ii) whether the
normal-state pseudogap generates the checkerboard charge ordering in the pseudogap phase?
In our recent work\cite{Zeng24}, the nature of the LDOS modulation of cuprate
superconductors in the SC-state for various kinds of impurities has been investigated
within the $T$-matrix approach, and the obtained results show that (i) the overall feature
of the LDOS modulation can be described qualitatively by taking into account the
quasiparticle scattering from a single impurity on the kinetic-energy-driven homogeneous
SC-state; (ii) the QSI scattering wave vectors ${\bf q}_{i}$ and the related QSI peak
dispersions are internally consistent within the octet scattering model. In this paper, we
study the distinct feature of the LDOS modulation in the pseudogap phase along with this
line. We identify explicitly that (i) the structure of the SC-state QSI patterns in the
LDOS modulation does not undergo any significant change as a function of temperature
through $T_{\rm c}$ to the pseudogap phase. Moreover, as in the SC-state\cite{Zeng24}, the
QSI peaks in the pseudogap phase are also dispersive with the dispersion of the QSI peaks
being particle-hole symmetric; (ii) the additional checkerboard peaks in the LDOS
modulation, which are suppressed strongly in the SC-state by the SC gap, emerge in the
pseudogap phase only. In particular, this checkerboard charge ordering is found to be
energy-independent. Our results therefore show that the occurrence of the unusual
checkerboard charge ordering is intrinsically connected with the opening of the
normal-state pseudogap.

This paper is organized as follows. The brief review of the interplay between the
pseudogap-state and superconductivity is presented in Section \ref{framework}, where in
the kinetic-energy-driven superconductivity, the normal-state pseudogap and SC gap
respectively originate from the homogeneous electron normal self-energy in the
particle-hole channel and anomalous self-energy in the particle-particle channel, and are
evaluated by taking into account the vertex correction. In Section \ref{QSI-LDOS}, we
start from the homogeneous electron propagator and the related {\it microscopic octet
scattering model} to derive LDOS in the pseudogap phase within the $T$-matrix approach,
where the $T$-matrix is calculated accurately for various kinds of a single impurity to
include all the quasiparticle excitations and scattering processes. The quantitative
characteristics of the LDOS modulation in the underdoped cuprate superconductors are
presented in Section \ref{Results}, where we show that the checkerboard charge-order wave
vector in the pseudogap phase obtained from the LDOS modulation is qualitatively
consistent with the results observed from the ARPES and STM/S experiments. Finally, we
give a brief summary in Sec. \ref{Summary}. In Appendix \ref{Derivation-of-T-matrix}, we
present the accurate calculation of the impurity-induced $T$-matrix in the pseudogap phase
in terms of the method of the inversion of matrix.

%\newpage

\section{Interplay between pseudogap-state and superconductivity}\label{framework}

\subsection{$t$-$J$ model with constrained electron}\label{model-constraint}

The key common feature in the crystal structure of cuprate superconductors is the
presence of the copper-oxide (CuO$_{2}$) plane\cite{Bednorz86,Wu87}, and then it seems
evident that the anomalous properties of cuprate superconductors are mainly governed
by these CuO$_{2}$ planes. In this case, as originally suggested by
Anderson\cite{Anderson87}, the essential physics of the doped CuO$_{2}$ plane can be
properly modeled with the square-lattice $t$-$J$ model,
\begin{equation}\label{tjham}
H=-\sum_{ll'\sigma}t_{ll'}C^{\dagger}_{l\sigma}C_{l'\sigma}
+\mu\sum_{l\sigma}C^{\dagger}_{l\sigma}
C_{l\sigma}+J\sum_{\langle ll'\rangle}{\bf S}_{l}\cdot {\bf S}_{l'},~~
\end{equation}
where $C^{\dagger}_{l\sigma}$ ($C_{l\sigma}$) creates (annihilates) an electron with
spin index $\sigma$ on lattice site $l$, ${\bf S}_{l}$ is a localized spin operator
with its components $S_{l}^{x}$, $S_{l}^{y}$, and $S_{l}^{z}$, and $\mu$ is the
chemical potential. This $t$-$J$ model (\ref{tjham}) contains two parts: the kinetic
energy part consists of the nearest-neighbor (NN) electron hopping term
$t_{ll'}=t_{\hat{\eta}}=t$ and the next NN electron hopping term
$t_{ll'}=t_{\hat{\tau}}=-t'$, while the magnetic energy part is described by a
Heisenberg term with the NN spin-spin antiferromagnetic (AF) exchange coupling $J$.
The summation $ll'$ is taken over all sites $l$, and for each $l$, over its NN sites
$\hat{\eta}$ and next NN sites $\hat{\tau}$, while the summation $\langle ll'\rangle$
is taken over all the NN pairs. For a convenience in the following discussions, the
NN spin-spin AF exchange coupling $J$ and the lattice constant $a$ of the square
lattice are set as the energy and length units, respectively. Although the values of
$J$, $t$, and $t'$ in cuprate superconductors are believed to vary somewhat from
compound to compound\cite{Damascelli03,Campuzano04,Fink07}, however, as a qualitative
discussion in this paper, the commonly used parameters are chosen as $t/J=2.5$ and
$t'/t=0.3$. In particular, when necessary to compare with the experimental data, we
set $J=100$meV.

The $t$-$J$ model (\ref{tjham}) is subjected to the on-site local constraint that
the double electron occupancy of a site by two electrons of opposite spins is not
allowed\cite{Yu92,Feng93,Zhang93,Lee06}. The physics of the charge-spin separation
appears in the fermion-spin theory\cite{Feng9404,Feng15} which starts from the
$t$-$J$ model (\ref{tjham}) and enforces the on-site local constraint of no double
electron occupancy by decoupling the physical electron into a spinful fermion and a
localized spin as,
\begin{eqnarray}\label{CSS}
C_{l\uparrow}=h^{\dagger}_{l\uparrow}S^{-}_{l}, ~~~~
C_{l\downarrow}=h^{\dagger}_{l\downarrow}S^{+}_{l},
\end{eqnarray}
where the $U(1)$ gauge invariant spinful fermion operator
$h^{\dagger}_{l\sigma}=e^{i\Phi_{l\sigma}}h^{\dagger}_{l}$
($h_{l\sigma}=e^{-i\Phi_{l\sigma}}h_{l}$) creates (annihilates) a charge carrier on
site $l$, and therefore describes the charge degree of freedom of the constrained
electron together with some effects of spin configuration rearrangements due to the
presence of the doped charge carrier itself, while $S^{+}_{l}$ ($S^{-}_{l}$) is the
$U(1)$ gauge invariant localized spin-raising (spin-lowering) operator, which describes
the spin degree of freedom of the constrained electron, and then the on-site local
constraint of the no double electron occupancy is always satisfied at each site.
Moreover, the collective mode from the spin degree of freedom of the constrained
electron can be interpreted as the spin excitation responsible for the spin dynamics of
the system, while the constrained electron as a result of the charge-spin recombination
of a charge carrier and a localized spin is responsible for the anomalous properties of
the electronic-state.

\subsection{Homogeneous electron propagator}
\label{IBSPS}

Based on the $t$-$J$ model (\ref{tjham}) in the fermion-spin representation, the
kinetic-energy-driven superconductivity has been
developed\cite{Feng15,Feng0306,Feng12,Feng15a,Li25}, where the coupling between charge
carriers directly from the kinetic energy in the fermion-spin theory (\ref{CSS})
description of the $t$-$J$ model (\ref{tjham}) by the exchange of the spin excitation
generates the d-wave charge-carrier pairing-state. The d-wave electron
pairs originated from the d-wave charge-carrier pairing-state are due to charge-spin
recombination\cite{Feng15a}, and these electron pairs condensation leads to form the
d-wave SC-state. The main features of the kinetic-energy-driven superconductivity can be
summarized as: (i) the mechanism is purely electronic without phonons; (ii) the mechanism
indicates that the strong electron correlation favors superconductivity, since the main
ingredient is identified into an electron pairing mechanism not involving the phonon, the
external degree of freedom, but {\it the spin excitation, the collective mode from the
internal spin degree of freedom of the constrained electron itself}. In other words, the
constrained electrons simultaneously act to glue and to be glued\cite{Schrieffer95,Xu23a};
(iii) the kinetic-energy-driven SC-state in a way is in turn strongly influenced by the
single-particle coherence, which leads to that $T_{\rm c}$ takes a dome-like shape with
the underdoped and overdoped regimes on each side of the optimal doping, where $T_{\rm c}$
reaches its maximum. Within the framework of the kinetic-energy-driven superconductivity,
the homogeneous electron propagator of the $t$-$J$ model (\ref{tjham}) has
been obtained explicitly as\cite{Feng15a,Li25},
\begin{subequations}\label{EDODGF}
\begin{eqnarray}
G({\bf k},\omega)&=&{1\over\omega-\varepsilon_{\bf k}-\Sigma_{\rm tot}({\bf k},\omega)},
~~~~~\label{EDGF}\\
\Im^{\dagger}({\bf k},\omega)&=&{W({\bf k},\omega)\over\omega-\varepsilon_{\bf k}
-\Sigma_{\rm tot}({\bf k},\omega)},~~~~~~~\label{EODGF}
\end{eqnarray}
\end{subequations}
with the electron energy dispersion in the tight-binding approximation
$\varepsilon_{\bf k}=-4t\gamma_{\bf k}+4t'\gamma_{\bf k}'+\mu$, where
$\gamma_{\bf k}=({\rm cos}k_{x}+{\rm cos} k_{y})/2$ and
$\gamma_{\bf k}'={\rm cos}k_{x}{\rm cos}k_{y}$, and the homogeneous electron total
self-energy $\Sigma_{\rm tot}({\bf k},\omega)$ and the function $W({\bf k},\omega)$ that
can be respectively expressed as,
\begin{subequations}\label{HTOTSE}
\begin{eqnarray}
\Sigma_{\rm tot}({\bf k},\omega)&=& \Sigma_{\rm ph}({\bf k},\omega)
+{|\Sigma_{\rm pp}({\bf k},\omega)|^{2}\over\omega +\varepsilon_{\bf k}
+\Sigma_{\rm ph}({\bf k},-\omega)},~~~~~\\
W({\bf k},\omega)&=&-{\Sigma_{\rm pp}({\bf k},\omega)\over\omega+\varepsilon_{\bf k}
+\Sigma_{\rm ph}({\bf k},-\omega)},
\end{eqnarray}
\end{subequations}
where $\Sigma_{\rm ph}({\bf k},\omega)$ is the homogeneous electron normal self-energy
in the particle-hole channel, and $\Sigma_{\rm pp}({\bf k},\omega)$ is the homogeneous
electron anomalous self-energy in the particle-particle channel, and have been obtained
by taking into account the vertex correction as\cite{Li25},
%\begin{widetext}
\begin{subequations}\label{E-N-AN-SE}
\begin{eqnarray}
\Sigma_{\rm ph}(&{\bf k}&,i\omega_{n})={1\over N^{2}}\sum_{{\bf p},{\bf p}'}
[V_{\rm ph}\Lambda_{{\bf p}+{\bf p}'+{\bf k}}]^{2}\nonumber\\
&\times& {1\over\beta}\sum_{ip_{m}}G({{\bf p}+{\bf k}},ip_{m}+i\omega_{n})
\Pi({\bf p},{\bf p}',ip_{m}),~~~~~~~~~~\label{E-N-SE}
\end{eqnarray}
\begin{eqnarray}
\Sigma_{\rm pp}(&{\bf k}&,i\omega_{n})={1\over N^{2}}\sum_{{\bf p},{\bf p}'}
[V_{\rm pp}\Lambda_{{\bf p}+{\bf p}'+{\bf k}}]^{2}\nonumber\\
&\times& {1\over \beta}\sum_{ip_{m}}\Im^{\dagger}({\bf p}+{\bf k},ip_{m}+i\omega_{n})
\Pi({\bf p},{\bf p}',ip_{m}),~~~~~~~~~~\label{E-AN-SE}
\end{eqnarray}
\end{subequations}
%\end{widetext}
respectively, with the bare vertex function
$\Lambda_{\bf k}=4t\gamma_{\bf k}-4t'\gamma_{\bf{k}}'$. The vertex correction
$V_{\rm ph}$ for the homogeneous electron normal self-energy, the vertex correction
$V_{\rm pp}$ for the homogeneous electron anomalous self-energy, together with the
spin bubble $\Pi({\bf p},{\bf p}',ip_{m})$ have been derived, and are given explicitly
in Ref. \onlinecite{Li25}. In low-temperatures, the sharp peaks emerge in
$\Sigma_{\rm ph}({\bf k},\omega)$ and $\Sigma_{\rm pp}({\bf k},\omega)$, which are
actually $\delta$-type functions. However, these $\delta$-type functions are broadened
by a small damping employed in the numerical calculation for a finite lattice. In this
paper, the calculation for $\Sigma_{\rm ph}(\bf k,\omega)$ and
$\Sigma_{\rm pp}({\bf k},\omega)$ is performed numerically on a $120\times 120$ lattice
in momentum space, where the infinitesimal $i0_{+}\rightarrow i\Gamma$ is replaced by a
small damping $\Gamma=0.05J$.

\subsection{Phase diagram}\label{Phase-diagram}

In the kinetic-energy-driven superconductivity\cite{Feng15,Feng0306,Feng12,Feng15a,Li25},
both the homogeneous electron normal and anomalous self-energies originate from the
coupling of the electrons with the same spin excitation, however, they describe
different parts of the coupling effect. The electron normal self-energy
$\Sigma_{\rm ph}({\bf k},\omega)$ describes the single-particle coherence, and its main
function is to renormalize the energy and lifetime of the electrons. In particular the
normal-state pseudogap is intimately related to the electron normal self-energy. To see
this intimate relation more clearly, we rewrite approximately the above electron normal
self-energy in Eq. (\ref{E-N-SE}) as\cite{Feng12,Feng15a,Li25},
\begin{equation}\label{ESE-NSPG}
\Sigma_{\rm ph}({\bf k},\omega) \approx {[2\bar{\Delta}_{\rm PG}({\bf k})]^{2}
\over\omega-\varepsilon_{\bf k}},
\end{equation}
where $\bar{\Delta}_{\rm PG}({\bf k})$ can be derived directly as,
\begin{equation}\label{NSPG}
\bar{\Delta}^{2}_{\rm PG}({\bf k})=-{1\over 4}\varepsilon_{\bf k}
\Sigma_{\rm ph}({\bf k},0),
\end{equation}
and is identified as being a region of the electron normal self-energy in which
$\bar{\Delta}_{\rm PG}$ anisotropically suppresses the electron density of states on
EFS\cite{Timusk99,Hufner08,Damascelli03,Campuzano04,Fink07,Hussey11,Vishik18}. In this
sense, $\bar{\Delta}_{\rm PG}$ can be identified as the momentum-dependent
normal-state pseudogap, with the normal-state pseudogap parameter
$\bar{\Delta}^{2}_{\rm PG}=(1/N)\sum_{\bf k}\bar{\Delta}^{2}_{\rm PG}({\bf k})$, and
then all the anomalous properties of the underdoped cuprate superconductors arise from
the opening of this normal-state pseudogap.
\begin{figure}[h!]
\includegraphics[scale=0.32]{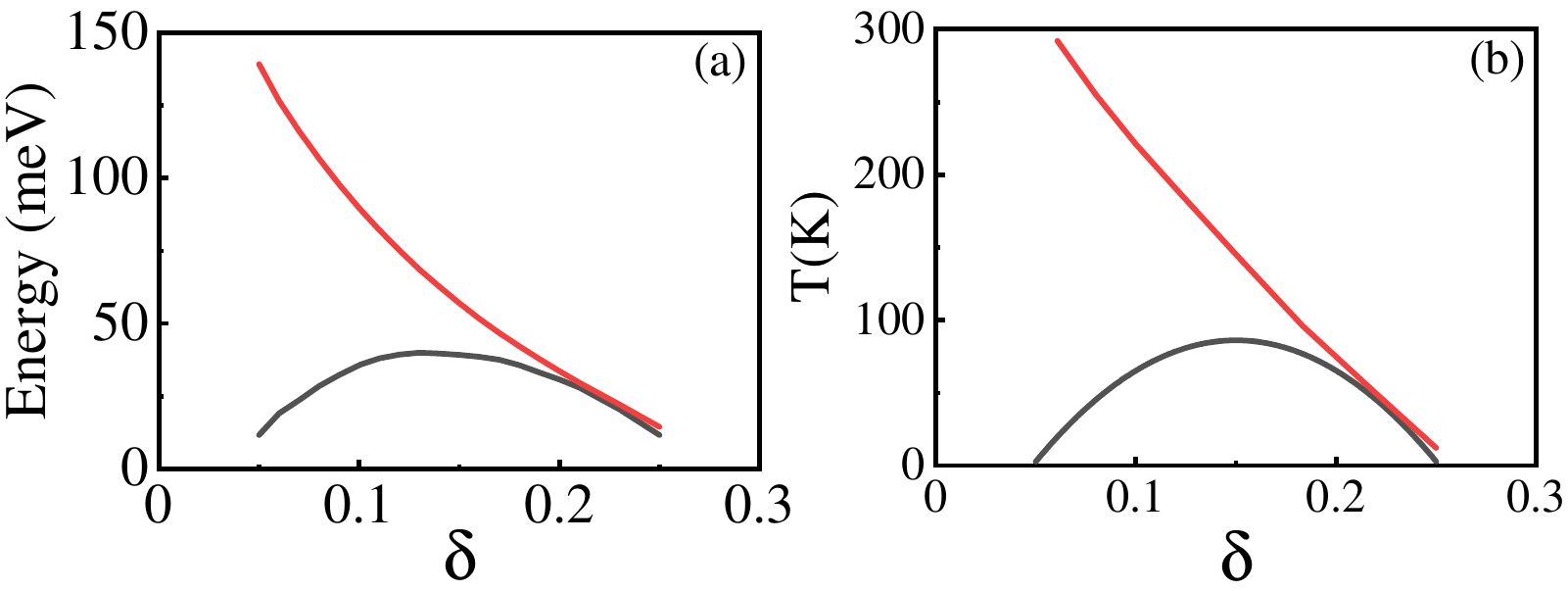}
\caption{(Color online) (a) The superconducting gap $2\bar{\Delta}$ (black-line) and
normal-state pseudogap $2\bar{\Delta}_{\rm PG}$ (red-line) as a function of doping
with temperature $T=0.002J$. (b) The superconducting transition temperature
$T_{\rm c}$ (black-line) and the normal-state pseudogap crossover temperature $T^{*}$
(red-line) as a function of doping.\label{EPG-T-doping}}
\end{figure}
However, the SC-state on the other hand is
characterized by the electron anomalous self-energy $\Sigma_{\rm pp}({\bf k},\omega)$,
which is defined as the energy- and momentum-dependent electron pair gap (then the SC
gap) $\Sigma_{\rm pp}({\bf k},\omega)=\bar{\Delta}_{\bf k}(\omega)$ in the the
single-particle excitation spectrum, and thus is corresponding to the energy for
breaking an electron pair\cite{Cooper56}. In the static limit, this SC gap
$\bar{\Delta}_{\bf k}(\omega)$ can be reduced as\cite{Feng15a,Li25},
\begin{eqnarray}\label{SC-gap}
\bar{\Delta}_{\bf k}(\omega)\mid_{\omega=0}=\bar{\Delta}_{\bf k}=\bar{\Delta}
\gamma^{\rm (d)}_{\bf k},
\end{eqnarray}
with the SC gap parameter $\bar{\Delta}$ and the d-wave factor
$\gamma^{\rm (d)}_{\bf k}=({\rm cos}k_{x}-{\rm cos} k_{y})/2$. The above results in
Eqs. (\ref{ESE-NSPG}) and (\ref{SC-gap}) therefore indicate that the pseudogap-state
coexists and competes with superconductivity below $T_{\rm c}$, although the
normal-state pseudogap bears no direct relation to superconductivity.

In the phase diagram of the cuprate superconductors, the crucial feature in the
underdoped regime is the intertwinement of the pseudogap-state and superconductivity.
To delve this intertwinement, the evolution of the pseudogap- and SC-states with
doping and temperature has been investigated\cite{Feng12,Feng15a,Li25}. For a better
understanding of the nature of the LDOS modulation in the pseudogap phase, the
results\cite{Li25} of the SC-gap 2$\bar{\Delta}$ (black-line) and normal-state
pseudogap 2$\bar{\Delta}_{\rm PG}$ (red-line) as a function of doping with temperature
$T=0.002J$ are replotted in Fig. \ref{EPG-T-doping}a, where the normal-state pseudogap
and SC gap diverge in the underdoped regime, i.e., the SC gap $\bar{\Delta}$ increases
smoothly with the increase of doping, and achieves its maximum at around the optimal
doping $\delta\approx 0.15$, while the normal-state pseudogap $\bar{\Delta}_{\rm PG}$
is notably larger than the SC gap $\bar{\Delta}$, and then it weakens as the optimal
doping is approached
\cite{Timusk99,Hufner08,Damascelli03,Campuzano04,Fink07,Hussey11,Vishik18,Drozdov18}.
However, in the overdoped regime, both the normal-state pseudogap and SC gap decrease
monotonically with the increase of doping, and converge to approximately same
magnitude in the strongly overdoped region, eventually going to zero together at the
end of the SC dome.

At a given doping concentration, the SC gap $\bar{\Delta}$ is identified with a SC
transition with a SC transition temperature $T_{\rm c}$, while the normal-state
pseudogap $\bar{\Delta}_{\rm PG}$ is defined as a crossover with a normal-state
pseudogap crossover temperature $T^{*}$. To see this doping dependence of $T_{\rm c}$
and $T^{*}$ more directly, the results\cite{Li25} of $T_{\rm c}$ (black-line) and
$T^{*}$ (red-line) as a function of doping are replotted in Fig. \ref{EPG-T-doping}b,
where in conformity with the doping dependence of $\bar{\Delta}$ and
$\bar{\Delta}_{\rm PG}$ shown in Fig. \ref{EPG-T-doping}a, (i) $T_{\rm c}$ is raised
monotonically upon the increase of doping in the underdoped regime, and achieves its
maximum at around the optimal doping, then $T_{\rm c}$ turns into a gradual decrease
in the overdoped regime; (ii) the pseudogap-state in the underdoped regime exhibits a
pseudogap at temperature $T^{*}$ large compared to $T_{\rm c}$, and then this $T^{*}$
decreases with the increase of doping, as opposed to $T_{\rm c}$. However, $T^{*}$
seems to degenerate with $T_{\rm c}$ in the strongly overdoped region, and goes
approximately to zero together with $T_{\rm c}$ at the end of the SC dome
\cite{Timusk99,Hufner08,Damascelli03,Campuzano04,Fink07,Hussey11,Vishik18,Drozdov18}.
This $T^{*}$ is actually a crossover temperature below which a novel electronic state
emerges, where the distinguished features are characterized by the presence of the
Fermi arcs\cite{Norman98,Shi08,Sassa11,Horio16,Loret18}, the static and fluctuating
ordering phenomena\cite{Comin16,Comin14,Gh12,Neto14,Campi15,Hash15}, the dramatic
change in the line-shape of the energy distribution
curve\cite{Hash15,Dessau91,Campuzano99,Lu01,Sato02,Borisenko03,Wei08,Loret17,DMou17},
and the kink in the quasiparticle
dispersion\cite{Bogdanov00,Kaminski01,Johnson01,Sato03,Lee09-1,He13}, etc., and are
correlated directly to the opening of the normal-state
pseudogap\cite{Timusk99,Hufner08,Damascelli03,Campuzano04,Fink07,Hussey11,Vishik18}.

\subsection{Reconstruction of constant energy contour}\label{EFSR-NS}

The essential feature of the low-energy quasiparticle excitation detected from ARPES
experiments can be analyzed theoretically in terms of the ARPES spectral
intensity\cite{Damascelli03,Campuzano04,Fink07},
\begin{equation}\label{QPES}
I({\bf k},\omega)\propto n_{\rm F}(\omega) A({\bf k},\omega),
\end{equation}
where $n_{\rm F}(\omega)$ is the fermion distribution, while the electron spectral
function $A({\bf k},\omega)$ is related directly with the imaginary part of the
 homogeneous electron diagonal propagator (\ref{EDGF}) as,
\begin{eqnarray}\label{ESF}
A({\bf k},\omega)&=& -{1\over\pi}{\rm Im}G({\bf k},\omega)\nonumber\\
&=& {1\over\pi}{\tilde{\Gamma}_{\bf k}(\omega)\over
[\omega-\varpi_{\bf k}(\omega)]^{2}+[\tilde{\Gamma}_{\bf k}(\omega)]^{2}},~~~
\end{eqnarray}
with the quasiparticle energy dispersion and quasiparticle scattering rate,
\begin{subequations}\label{QPED-QPSR}
\begin{eqnarray}
\varpi_{\bf k}(\omega)&=&\varepsilon_{\bf k}
-{\rm Re}\Sigma_{\rm tot}({\bf k},\omega),\label{QPED}\\
\tilde{\Gamma}_{\bf k}(\omega)&=& {\rm Im}\Sigma_{\rm tot}({\bf k},\omega),
~~~~~~\label{QPSR}
\end{eqnarray}
\end{subequations}
in the SC-state, respectively, where ${\rm Re}\Sigma_{\rm tot}({\bf k},\omega)$ and
${\rm Im}\Sigma_{\rm tot}({\bf k},\omega)$ are the real and imaginary parts of the
homogeneous electron total self-energy $\Sigma_{\rm tot}({\bf k},\omega)$. In
particular, when the SC gap parameter $\bar{\Delta}=0$ in the pseudogap phase above
$T_{\rm c}$, these quasiparticle energy dispersion and quasiparticle scattering rate
in Eq. (\ref{QPED-QPSR}) in the SC-state are reduced as,
\begin{subequations}\label{QPED-QPSR-NS}
\begin{eqnarray}
\varpi_{\bf k}(\omega)&=&\varepsilon_{\bf k}-{\rm Re}\Sigma_{\rm ph}({\bf k},\omega),
\label{QPED-NS}\\
\tilde{\Gamma}_{\bf k}(\omega)&=& {\rm Im}\Sigma_{\rm ph}({\bf k},\omega)\approx \pi
{[2\bar{\Delta}_{\rm PG}({\bf k})]^{2}\delta(\omega-\varepsilon_{\bf k}}),~~~~~~~~
\label{QPSR-NS}
\end{eqnarray}
\end{subequations}
in the pseudogap phase, respectively, with the real part
${\rm Re}\Sigma_{\rm ph}({\bf k},\omega)$ and the imaginary part
${\rm Im}\Sigma_{\rm ph}({\bf k},\omega)$ of the homogeneous electron normal self-energy,
respectively. The above results in Eqs. (\ref{QPSR}) and (\ref{QPSR-NS}) therefore show
clearly whether in the SC-state or in the normal-state, the quasiparticle scattering
rate is intrinsically correlated to the normal-state pseudogap, which is consistent with
the experimental observations\cite{Matt15}.

In the previous discussions, both the experimental
observations\cite{Chatterjee06,McElroy06,Chatterjee07,He14,Restrepo23,Shah25} and
theoretical studies\cite{Gao19,Liu21,Zeng22} indicate that in the underdoped regime, the
constant energy contour in the SC-state at the case for a finite binding-energy is
truncated into the Fermi arcs centered at around the nodal region, where the notation
{\it Fermi arcs} on the constant energy contour has been used even for a finite
binding-energy. Moreover, these experimental observations and theoretical studies
also show that the Fermi arcs are not strongly peaked intensities at around the nodes,
however, the highest intensities are located at around the tips of the Fermi arcs. These
tips of the Fermi arcs connected by the scattering wave vectors ${\bf q}_{i}$ then
construct an {\it octet scattering
model}\cite{Chatterjee06,McElroy06,Chatterjee07,He14,Restrepo23,Shah25}, which is a
basic scattering model for the interpretation of the SC-state QSI in cuprate
superconductors
\cite{Pan01,Fischer07,Yin21,Zeng22,Zeng23,Pan00,Hoffman02,McElroy03,Kohsaka08,Hanaguri09,Vishik09}.
Since the SC- and pseudogap-states coexist below $T_{\rm c}$ and the
quasiparticle scattering rate is proportional to the normal-state pseudogap, it has been
thus shown that the formation of the Fermi arcs and the related octet scattering model
due to the redistribution of the spectral weight on the constant energy contour for a
finite binding-energy is directly linked with the opening of the highly anisotropic
momentum-dependence of the normal-state pseudogap.

\begin{figure}[h!]
\centering
\includegraphics[scale=0.30]{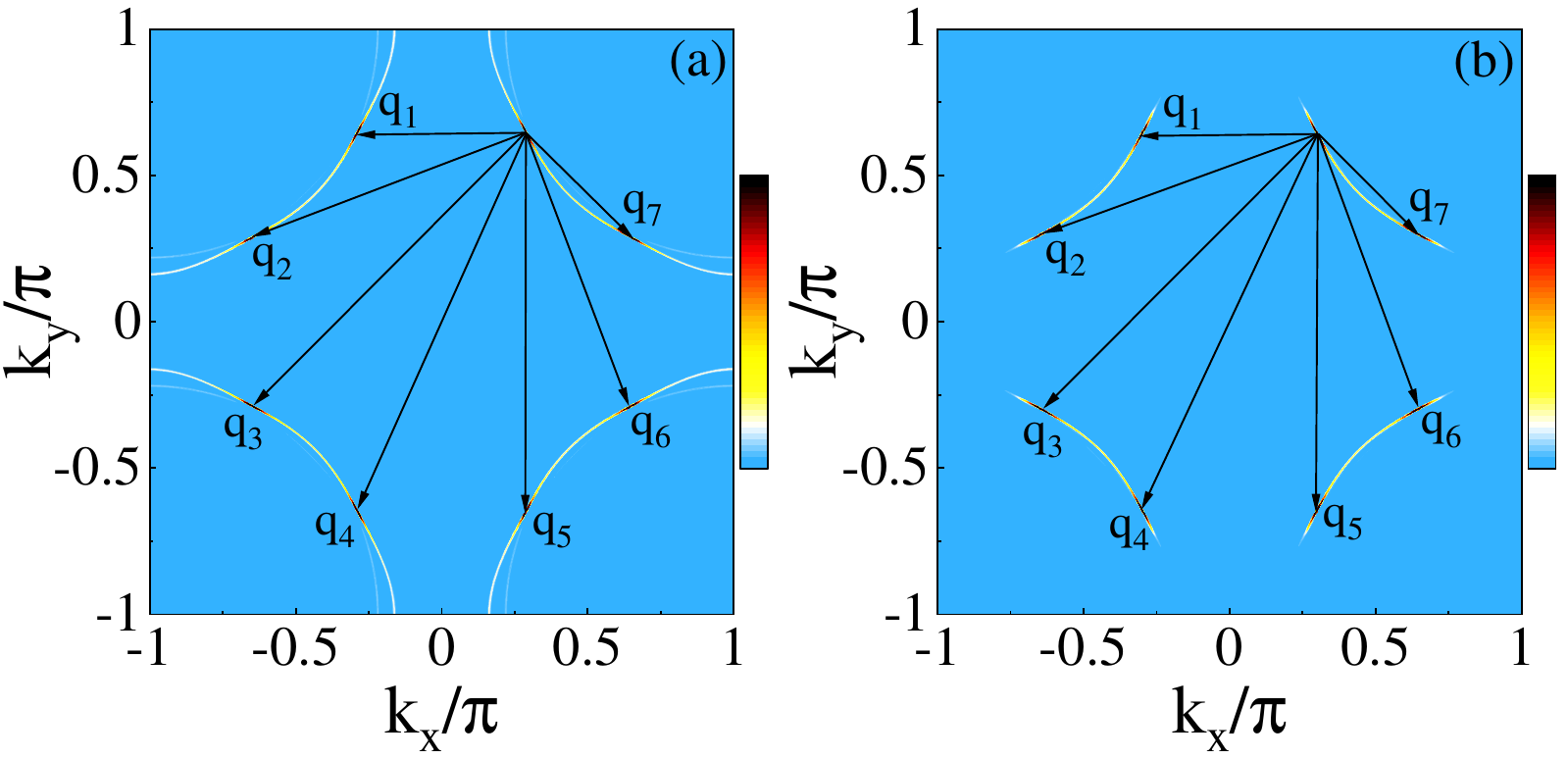}
\caption{(Color online) The constant energy contour map in the $[k_{x},k_{y}]$ plane at
doping $\delta=0.09$ for the binding-energy $\omega=0.06J$ in (a) the pseudogap phase
with temperature $T=0.07J$ and (b) the superconducting-state with temperature $T=0.002J$.
The eight tips of the Fermi arcs determine the scattering within the octet scattering
model, while the scattering wave vectors which connect the tips of the Fermi arcs are
shown as arrows labeled by the designation of each scattering wave vector ${\bf q}_{i}$.
\label{CEC-maps}}
\end{figure}

Very recently, we\cite{Li25} have studied the exotic features of the low-energy
electronic structure of cuprate superconductors in the pseudogap phase, where we found
that the characteristic of the octet scattering model in the SC-state remain unchanged
upon passing above $T_{\rm c}$ to the pseudogap phase, in agreement with the experimental
observations\cite{Chatterjee06,McElroy06,Chatterjee07,He14,Restrepo23,Shah25}. For a
convenience in the following discussion of the LDOS modulation in the pseudogap phase,
the constant energy contours in the $[k_{x},k_{y}]$ plane at doping $\delta=0.09$ for the
binding-energy $\omega=0.06J$ in (a) the pseudogap phase with temperature $T=0.07J$ and
(b) the SC-state with temperature $T=0.002J$ are replotted in Fig. \ref{CEC-maps}.
Expectedly, as in the case of the SC-state\cite{Gao19,Liu21,Zeng22}, (i) the structure of
the constant energy contour intensity maps and the dispersions of the seven scattering wave
vectors ${\bf q}_{i}(\omega)$ $[i=1,2,...,7]$ are essentially the same both in the SC-state
and pseudogap phase.
In particular, the tips of the Fermi arcs connected by the scattering
wave vectors ${\bf q}_{i}(\omega)$ in the pseudogap phase still construct an {\it octet
scattering model}; (ii) all seven scattering wave vectors ${\bf q}_{i}(\omega)$ retain
their particle-hole symmetry ${\bf q}_{i}(\omega)={\bf q}_{i}(-\omega)$ in the pseudogap
phase, then the LDOS modulations occur in the same energy range, and emanate from the same
constant energy contour in momentum-space as those observed in the SC-state.
However, there is a
substantial difference between the pseudogap phase and SC-state at around the antinodal
region, namely, the spectral weight at around the antinodal region in the SC-state is
gapped out completely by both the SC gap and normal-state pseudogap, while the spectral
weight at around the antinodal region in the pseudogap phase is suppressed partially by the
normal-state pseudogap only, which leads to that small residual spectral-weight exists at
around the antinodal region in the pseudogap phase. This small residual spectral-weight may
induce the unusual charge ordering in the pseudogap phase, and we will discuss this issue
towards in Sec. \ref{Results}.

%\newpage

\section{Local density of states}\label{QSI-LDOS}

\subsection{Local density of states in superconducting-state}\label{LDOS-SC}

In the recent work\cite{Zeng24}, LDOS in the SC-state has been derived for various kinds
of impurities. For the discussion of the LDOS modulation in the pseudogap phase and then
comparison it with the LDOS modulation in the SC-state, here the main formalism of LDOS
in the SC-state and the related $T$-matrix are summarized briefly. In the presence of
impurities with the impurity-scattering potential $V(\bm{r})$, the homogeneous electron
propagator (\ref{EDODGF}) in the SC-state is dressed by the impurity scattering, while
this dressed electron propagator has been obtained within the $T$-matrix approach
as\cite{Zeng24},
\begin{eqnarray}\label{Green-Fun-k}
\tilde{G}_{I}(\bm{k},\bm{k}',\omega)=\tilde{G}(\bm{k},\omega)\delta_{\bm{k},\bm{k}'}
+\tilde{G}(\bm{k},\omega)\tilde{T}_{\bm{k}\bm{k}'}(\omega)\tilde{G}(\bm{k}',\omega),
~~~~~
\end{eqnarray}
where $\tilde{G}(\bm{k},\omega)$ is the homogeneous electron propagator (\ref{EDODGF})
in the Nambu representation, and the $T$-matrix $\tilde{T}_{\bm{k}\bm{k}'}(\omega)$
satisfies the following iteration equation,
\begin{eqnarray}\label{T-matrix-equation-1}
\tilde{T}_{\bm{k}\bm{k}'}(\omega)={1\over N}V_{\bm{k}\bm{k}'}\tau_{3}
+{1\over N}\sum_{\bm{k}_{2}}
V_{\bm{k}\bm{k}_{2}}\tau_{3}\tilde{G}(\bm{k}_{2},\omega)
\tilde{T}_{\bm{k}_{2}\bm{k}'}(\omega), ~~~~~
\end{eqnarray}
with the impurity-scattering potential in momentum-space,
\begin{eqnarray}
V_{\bm{k}_{1}\bm{k}_{2}}=V(\bm{k}_{1}-\bm{k}_{2})=\sum_{\bm{r}}
e^{-i(\bm{k}_{1}-\bm{k}_{2})\cdot\bm{r}}V(\bm{r}).
\end{eqnarray}
In this case, the corresponding form of the dressed electron propagator in real-space can
be derived directly in terms of the Fourier transform as,
\begin{eqnarray}\label{Green-Fun-k-1}
\tilde{G}_{I}(&\bm{r}&,\bm{r}',\omega)=\tilde{G}(\bm{r}-\bm{r}',\omega)\nonumber\\
&+&{1\over N}\sum_{\bm{k},\bm{k}'}e^{i\bm{k}\cdot\bm{r}-i\bm{k}'\cdot\bm{r}'}
\tilde{G}(\bm{k},\omega)\tilde{T}_{\bm{k}\bm{k}'}(\omega)\tilde{G}(\bm{k}',\omega).
~~~~~~
\end{eqnarray}
The above results in Eqs. (\ref{Green-Fun-k}) and (\ref{Green-Fun-k-1}) thus indicate
unambiguously that the nature of the LDOS modulation is governed by both the homogeneous
electron propagator (then the constant energy contour and the related octet scattering
model), which reflects the effect of the strong electron correlation on the LDOS
modulation, and the impurity-generated $T$-matrix, which incorporates both the effects
of the strong electron correlation and impurity-scattering on the LDOS modulation. The
above results in Eqs. (\ref{Green-Fun-k}) and (\ref{Green-Fun-k-1}) also show that in
additional to the homogeneous electron propagator, the more accurate result of the
$T$-matrix should be obtained in the proper discussion of the LDOS modulation. In this
case, we\cite{Zeng24} have developed a powerful method of the inversion of matrix, and
then the above impurity-generated $T$-matrix in Eq. (\ref{T-matrix-equation-1}) has been
derived accurately as,
\begin{equation}\label{Tmat-Expression}
\tilde{T}(\omega)=\bar{V}\otimes\tau_{0}{1\over 1-\bar{M}}\hat{I}_{v}\otimes\tau_{3},
\end{equation}
with a unit matrix in momentum-space $\hat{I}_{v}$, and the matrices $\bar{V}$ and
$\bar{M}$ that have been given explicitly in Ref. \onlinecite{Zeng24}. It should be
emphasized that in this approach of the inversion of matrix, the derivation of the
$T$-matrix consists of all the quasiparticle excitations and scattering processes.

The real-space LDOS modulation $\delta\rho(\bm{r},\omega)$ now can be evaluated
directly from the imaginary part of the diagonal matrix elements of the
impurity-dressed electron propagator (\ref{Green-Fun-k-1}) as,
\begin{eqnarray}\label{Rhoq-R}
&&\delta\rho(\bm{r},\omega)=\rho(\bm{r},\omega)-\rho_{0}(\omega)\nonumber\\
&=&-{2\over\pi}{\rm Im}\Big[{1\over N}\sum_{\bm{k}\bm{k}'}
e^{i(\bm{k}-\bm{k}')\cdot\bm{r}}\tilde{G}(\bm{k},\omega)
\tilde{T}_{\bm{k}\bm{k}'}(\omega)\tilde{G}(\bm{k}',\omega)\Big]_{11},~~~~~~
\end{eqnarray}
with the density of states in the case of the absence of impurities $\rho_{0}(\omega)$.
The momentum-space LDOS modulation $\delta\rho(\bm{q},\omega)$ is easily obtained
from the above real-space LDOS modulation $\delta\rho(\bm{r},\omega)$ in
Eq. (\ref{Rhoq-R}) by using the Fourier-transform as,
\begin{equation}\label{Rhoq}
\delta\rho(\bm{q},\omega)=\text{Re}[\delta\rho(\bm{q},\omega)]
+ i\text{Im}[\delta\rho(\bm{q},\omega)],
\end{equation}
where $\text{Re}[\delta\rho(\bm{q},\omega)]$ and $\text{Im}[\delta\rho(\bm{q},\omega)]$
are respectively the corresponding real and imaginary parts of the LDOS modulation
$\delta\rho(\bm{q},\omega)$, and can be expressed explicitly as,
\begin{subequations}
\begin{eqnarray}
\text{Re}[\delta\rho(\bm{q},\omega)]&=&-\frac{1}{\pi}\text{Im}
[\tilde{A}_{\rm I}(\bm{q},\omega)+\tilde{A}_{\rm I}(-\bm{q},\omega)]_{11},
\label{Rhoq-Re}\\
\text{Im}[\delta\rho(\bm{q},\omega)]&=&-\frac{1}{\pi}\text{Re}
[\tilde{A}_{\rm I}(\bm{q},\omega)-\tilde{A}_{\rm I}(-\bm{q},\omega)]_{11}, ~~~~~
\label{Rhoq-Im}
\end{eqnarray}
\end{subequations}
with the related spectral function,
\begin{equation}
\tilde{A}_{\rm I}(\bm{q},\omega)=\sum_{\bm{k}}\tilde{G}(\bm{k},\omega)
\tilde{T}(\bm{k},\bm{k}+\bm{q},\omega)\tilde{G}(\bm{k}+\bm{q},\omega).~~~~
\end{equation}

\subsection{Local density of states in pseudogap phase}\label{LDOS-NSPH}

In the pseudogap phase, where the SC gap $\bar{\Delta}=0$, the homogeneous electron
propagator (\ref{EDODGF}) in the SC-state is reduced as,
\begin{eqnarray}\label{normal-propagator}
G({\bf k},\omega)={1\over\omega-\varepsilon_{\bf k}-\Sigma_{\rm ph}({\bf k},\omega)},
\end{eqnarray}
and then the impurity-dressed electron propagator (\ref{Green-Fun-k}) in the
SC-state and the related $T$-matrix iteration equation (\ref{T-matrix-equation-1})
are respectively reduced as,
\begin{subequations}
\begin{eqnarray}
G_{I}(\bm{k},\bm{k}',\omega) &=& G(\bm{k},\omega)\delta_{\bm{k},\bm{k}'}
+G(\bm{k},\omega)T_{\bm{k}\bm{k}'}(\omega)G(\bm{k}',\omega),~~~
\label{Normal-Green-Fun-Eq}\nonumber\\
~~~~\\
T_{\bm{k}\bm{k}'}(\omega) &=& \bar{V}_{\bm{k}\bm{k}'}+\sum_{\bm{k}_{1}}
\bar{V}_{\bm{k}\bm{k}_{1}}G(\bm{k}_{1},\omega)T_{\bm{k}_{1}\bm{k}'}(\omega).
\label{norman-T-matrix-equation-1}
\end{eqnarray}
\end{subequations}
As in the case of the SC-state $T$-matrix (\ref{Tmat-Expression}) mentioned in the
above subsection \ref{LDOS-SC}, the $T$-matrix in the pseudogap phase can be also
derived accurately by making use of the approach of the inversion of matrix as [see
Appendix \ref{Derivation-of-T-matrix}],
\begin{equation}\label{normal-Tmat-Expression}
T(\omega)=\bar{V}{1\over 1-\bar{M}}.
\end{equation}
With the help of the Fourier-transform, the impurity-dressed electron propagator
$G_{I}(\bm{r},\bm{r}',\omega)$ in real-space now can be derived straightforwardly from
the impurity-dressed electron propagator $G_{I}(\bm{k},\bm{k}',\omega)$ in
Eq. (\ref{Normal-Green-Fun-Eq}) in momentum-space as,
\begin{eqnarray}\label{normal-Green-Fun-k-1}
G_{I}(\bm{r},\bm{r}',\omega)&=&G(\bm{r}-\bm{r}',\omega)
+{1\over N}\sum_{\bm{k},\bm{k}'}e^{i\bm{k}\cdot\bm{r}-i\bm{k}'\cdot\bm{r}'}
G(\bm{k},\omega)\nonumber\\
&\times& T_{\bm{k}\bm{k}'}(\omega)G(\bm{k}',\omega),
\end{eqnarray}
and then the real-space LDOS modulation $\delta\rho(\bm{r},\omega)$ can be obtain from
the imaginary part of the above impurity-dressed electron propagator
(\ref{normal-Green-Fun-k-1}) as,
\begin{eqnarray}\label{normal-Rhoq-R}
\delta\rho(\bm{r},\omega) &=& -{1\over\pi}{\rm Im}\Big[{1\over N}\sum_{\bm{k}\bm{k}'}
e^{i(\bm{k}-\bm{k}')\cdot\bm{r}}G(\bm{k},\omega)\nonumber\\
&\times& T_{\bm{k}\bm{k}'}(\omega)G(\bm{k}',\omega)\Big].~~~~
\end{eqnarray}
In this case, the corresponding momentum-space LDOS modulation
$\delta\rho(\bm{q},\omega)$ can be evaluated directly from the above real-space LDOS
modulation $\delta\rho(\bm{r},\omega)$ in terms of the Fourier-transform as,
\begin{equation}\label{RhoqN}
\delta\rho(\bm{q},\omega)=\text{Re}\delta\rho(\bm{q},\omega)
+ i \text{Im}\delta\rho(\bm{q},\omega),
\end{equation}
where $\text{Re}\delta\rho(\bm{q},\omega)$ and $\text{Im}\delta\rho(\bm{q},\omega)$
are given as,
\begin{subequations}\label{rhoqn-RIM-Part}
\begin{eqnarray}
\text{Re}\delta\rho(\bm{q},\omega)&=&  -\frac{1}{\pi} \text{Im}
[A_{\rm I}(\bm{q},\omega) + A_{\rm I}(-\bm{q},\omega) ], \\
\text{Im}\delta\rho(\bm{q},\omega) &=&-\frac{1}{\pi} \text{Re}
[A_{\rm I}(\bm{q},\omega)- A_{\rm I}(-\bm{q},\omega)],
\end{eqnarray}
\end{subequations}
respectively, with the related spectral function,
\begin{equation}
A_{\rm I}(\bm{q},\omega)=\sum_{\bm{k}}G(\bm{k},\omega)T(\bm{k},\bm{k}+\bm{q},\omega)
G(\bm{k}+\bm{q},\omega).
\end{equation}
According to the above real and imaginary parts of the momentum-space LDOS modulation
in Eq. (\ref{rhoqn-RIM-Part}), the weight $|\delta\rho(\bm{q},\omega)|$ and the phase
$\phi(\bm{q},\omega)$ of the LDOS modulation spectrum are obtained explicitly as,
\begin{subequations}
\begin{eqnarray}
|\delta\rho(\bm{q},\omega)| &=& \sqrt{\text{Re}\delta\rho(\bm{q},\omega)^{2}
+ \text{Im}\delta\rho(\bm{q},\omega)^2}, \\
\phi(\bm{q},\omega) &=& \arctan \left ( \frac{\text{Im}\delta\rho(\bm{q},\omega)}
{\text{Re}\delta\rho(\bm{q},\omega)} \right )
+ \pi\theta(-\text{Re}\delta\rho(\bm{q},\omega)),~~\nonumber\\
~~~~
\end{eqnarray}
\end{subequations}
respectively. The characteristic feature in the momentum-space LDOS modulation
spectrum $\delta\rho(\bm{q},\omega)$ is governed by the
peaks\cite{Vershinin04,Hanaguri04,McElroy05,Wise08,Lee09,Schmidt11,Fujita14}, while the
peaks for a given energy $\omega$ is corresponding to the momentum-space highest joint
density of states at that energy $\omega$, and then $\bm{q}(\omega)$ of the peaks in the
LDOS modulation spectrum can be straightforwardly connected to the quasiparticle
dispersion.

%\newpage

\section{Quantitative Characteristics}\label{Results}

In this section, we discuss the quantitative characteristics of the LDOS modulation
in the underdoped cuprate superconductors generated by various kinds of a single
impurity to shed light on the nature of the quasiparticle excitation in the pseudogap
phase and of its similarity and difference with the corresponding one in the SC-state.
As in our recent studies\cite{Zeng24}, the strength of the impurity scattering potential
in the following discussion is set to be positive to refrain from the quantum resonant
state, since the $T$-matrix approach is not valid for the case of the quantum resonant
state generated by the impurity scattering potential with the negative strength.

\begin{figure*}[t!]
%\begin{figure}[h!]
\centering
\includegraphics[scale=0.45]{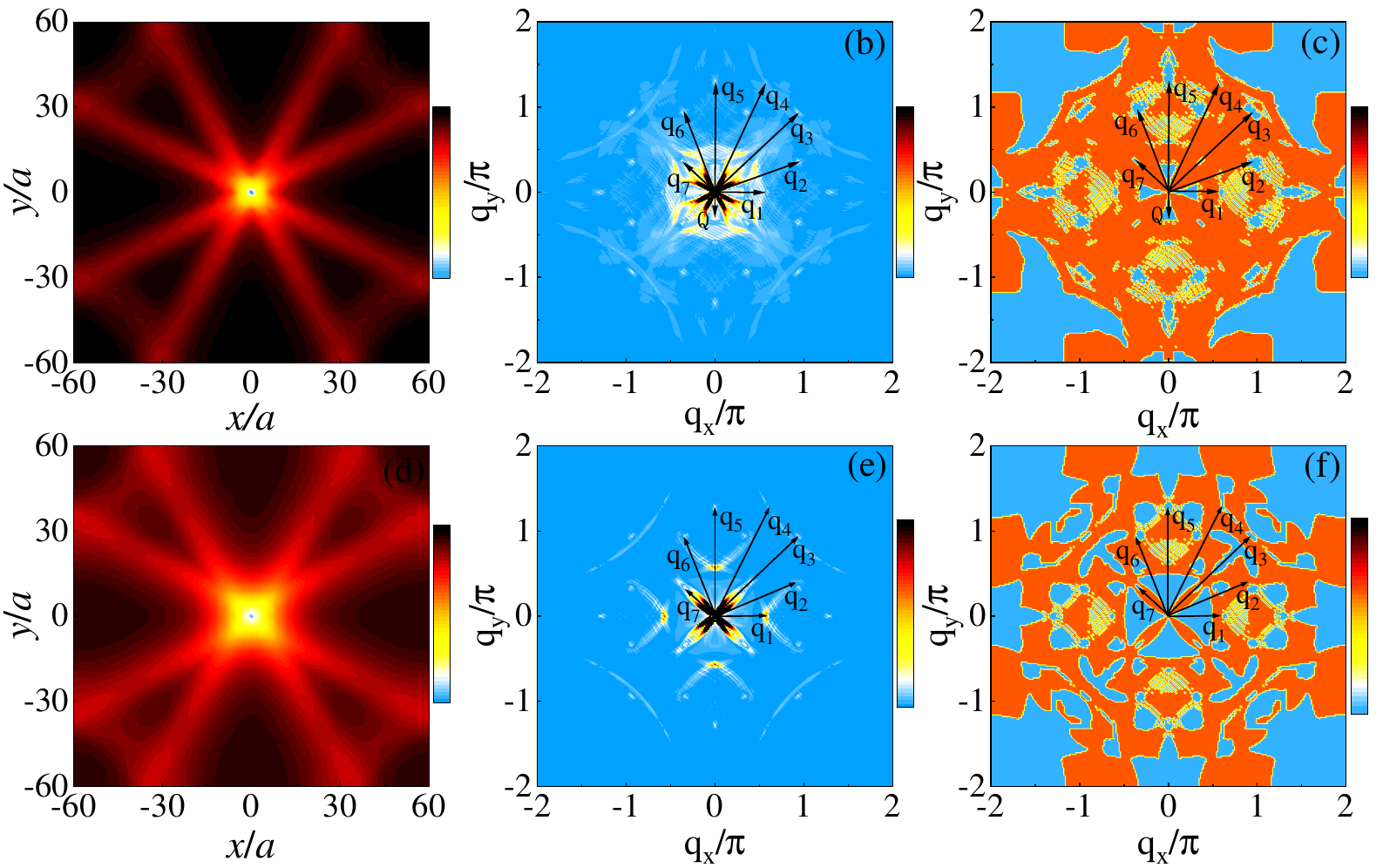}
\caption{(Color online) Upper panel: the intensity maps of (a) the local density
of states in real-space, (b) the amplitude of the local density of states in
momentum-space, and (c) the phase of the momentum-space local density of states in
the pseudogap phase for an in-plane single impurity at doping $\delta=0.09$ and the
binding-energy $\omega=0.06J$ with temperature $T=0.07J$ for the impurity-scattering
strength $V_{s}=1.0J$ and screening length $L=5.0a$. Lower panel: the corresponding
intensity maps of (d) the local density of states in real-space, (e) the amplitude
of the local density of states in momentum-space, and (f) the phase of the
momentum-space local density of states in the superconducting-state with temperature
$T=0.002J$.\label{QSI-in-plane}}
%\end{figure}
\end{figure*}

The crystal structure of cuprate superconductors is a stacking of the
common CuO$_{2}$ planes separated by insulating layers\cite{Bednorz86,Wu87}, the
impurity distribution accompanied with different types of the doping processes are
quite different\cite{Hussey02,Balatsky06,Alloul09}. In particular, impurities which
substitute for Cu in the CuO$_{2}$ plane turn out to be strong scatterers of the
electronic state in the CuO$_{2}$ plane, giving rise to a major modification of the
electronic structure. The impurity-scattering potential for an in-plane single
impurity located at the lattice site $\bm{r}=0$ can be modeled in real-space by a
screened Coulomb potential with the screening length
as\cite{Hussey02,Balatsky06,Alloul09},
\begin{eqnarray}\label{VPOT-RS}
V(\bm{r})={V_{s}\over |\bm{r}|}e^{-|\bm{r}|/L},
\end{eqnarray}
where $V_{s}$ is the in-plane impurity-scattering strength, $L$ is the screening
length, while its Fourier-transformed form can be derived straightforwardly as,
\begin{equation}\label{VPOT-1}
V_{\bm{k}\bm{k}'}={2\pi V_{s}\over \sqrt{(\bm{k}-\bm{k}')^{2}+1/L^2}}.
\end{equation}
In the following discussion, the in-plane impurity-scattering strength $V_{s}$ and
its screening length $L$ are chosen as $V_{s}=1.0J$ and $L=5.0a$, respectively.

To characterize the LDOS modulation in the pseudogap phase, we have performed
a series of calculations for LDOS both in real- and momentum-spaces for an
in-plane single impurity in the {\it pseudogap phase}, and the obtained results of
the intensity maps of (a) the LDOS modulation in real-space, (b) the amplitude of
the LDOS modulation in momentum-space, and (c) the phase of the momentum-space LDOS
modulation at doping $\delta=0.09$ and the binding-energy $\omega=0.06J$ with
temperature $T=0.07J$ for the impurity-scattering strength $V_{s}=1.0J$ and
screening length $L=5.0a$ are plotted in the upper panel of Fig. \ref{QSI-in-plane}.
For a clear comparison, the corresponding results\cite{Zeng24} of the intensity
maps of (d) the LDOS modulation in real-space, (e) the amplitude of the LDOS
modulation in momentum-space, and (f) the phase of the momentum-space LDOS
modulation in the SC-state with temperature $T=0.002J$ are replotted in the lower
panel of Fig. \ref{QSI-in-plane}. The main features of the obtained results can
be summarized as:\\
(i) The global feature of the real-space LDOS modulation in the pseudogap phase
is similar to that in the SC-state (see Fig. \ref{QSI-in-plane}a and
Fig. \ref{QSI-in-plane}d), i.e., the real-space LDOS modulation in the pseudogap
phase presents a similar ripple pattern as that in the SC-state. However, the
weight of the LDOS modulation in momentum-space centered at around the nodal
region is effectively reduced in the pseudogap phase, which in turn weakens the
LDOS modulation in real-space along the diagonal direction, leading to that the
intensity of the ripples is decreased, and the weight of the LDOS modulation
connected two ripples along the diagonal direction vanishes in the pseudogap phase.
\begin{figure}[h!]
\centering
\includegraphics[scale=0.32]{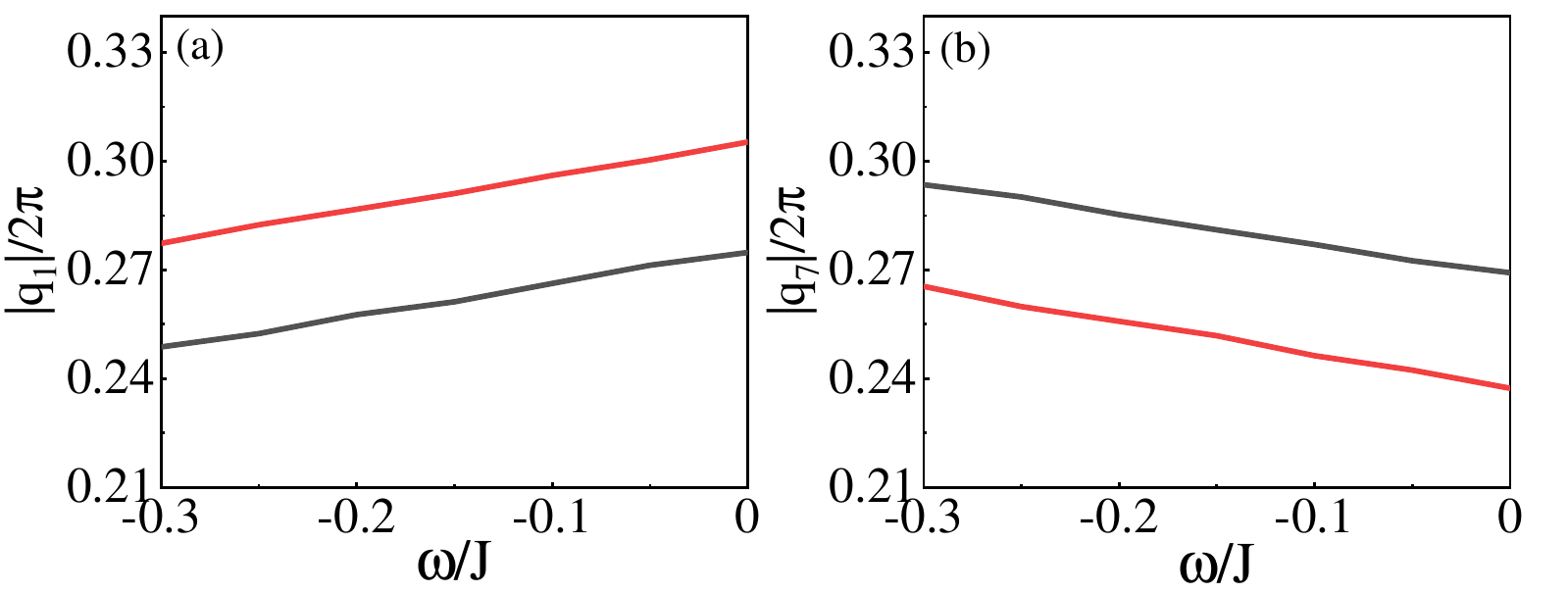}
\caption{(Color online) The evolution of the scattering wave vectors (a)
${\bf q}_{1}$ and (b) ${\bf q}_{7}$ with energy in the pseudogap phase for an
in-plane single impurity at doping $\delta=0.09$ (red-line) and doping
$\delta=0.12$ (black-line) with temperature $T=0.07J$ for the impurity-scattering
strength $V_{s}=1.0J$ and screening length $L=5.0a$. \label{LDOS-MS-E}}
\end{figure}
Furthermore, the fuzzy region in the momentum-space LDOS modulation pattern with
small momenta along the antinodal direction appears in the pseudogap phase, while
the small weight in this fuzzy region in the pseudogap phase is suppressed
strongly in the SC-state by the SC gap, leading to a more clearer SC-state QSI
pattern in the momentum-space LDOS modulation;\\
(ii) The SC-state QSI peaks in the momentum-space LDOS modulation are located
exactly at the discrete spots\cite{Zeng24} (see Fig. \ref{QSI-in-plane}e), where
these discrete spots identified as regions of the high joint density of states are
accommodated at around the scattering wave vectors ${\bf q}_{i}$ with
$i=1,2,\cdots,7$ obeying the {\it octet scattering model} shown in
Fig. \ref{CEC-maps}b. Moreover, all these QSI signatures detectable in the SC-state
survive virtually unchanged into the pseudogap phase (see Fig. \ref{QSI-in-plane}b),
i.e., the QSI octet phenomenology in the
SC-state\cite{Pan00,Hoffman02,McElroy03,Kohsaka08,Hanaguri09,Vishik09} exists in the
pseudogap phase\cite{Vershinin04,Hanaguri04,McElroy05,Wise08,Lee09,Schmidt11,Fujita14},
then the multiple ordered electronic-states, which emerge in the pseudogap phase
above $T_{\rm c}$, persist into the SC-state below $T_{\rm c}$ and coexist and compete
with the SC-state. As a natural consequence, all seven scattering wave vectors
${\bf q}_{i}(\omega)$, which are dispersive in the SC-state, remain dispersive into
the pseudogap phase. To characterize typical QSI peak dispersions in the pseudogap
phase more clearly, we plot the energy dependence of the QSI peaks in the pseudogap
phase for the scattering wave vectors (a) ${\bf q}_{1}$ and (b) ${\bf q}_{7}$ at
doping $\delta=0.09$ (red-line) and doping $\delta=0.12$ (black-line) with
temperature $T=0.07J$ for an in-plane single impurity for the impurity-scattering
strength $V_{s}=1.0J$ and screening length $L=5.0a$ in Fig. \ref{LDOS-MS-E}, where
for a given doping concentration, the peak at the scattering wave vector
${\bf q}_{1}$ evolves very differently with energy than the peak at the scattering
wave vector ${\bf q}_{7}$, and their dispersions have opposite sign, i.e., the length
of ${\bf q}_{1}$ becomes shorter, whereas the length of ${\bf q}_{7}$ becomes longer,
as energy $|\omega|$ is increased. However, with the increase of the doping
concentration, ${\bf q}_{1}$ decreases systematically in length, while ${\bf q}_{7}$
moves to higher values, which are expected since the distance between the parallel
tips of the Fermi arcs moves closer together, while the distance between the diagonal
tips of the Fermi arcs increases\cite{Zeng22,Gao19}.
The results of the evolution of the QSI peaks with energy and doping in
Fig. \ref{LDOS-MS-E} therefore further confirm that as in the case of the
SC-state\cite{Pan00,Hoffman02,McElroy03,Kohsaka08,Hanaguri09,Vishik09}, the typical
QSI peak dispersions in the pseudogap phase are also internally consistent within the
octet scattering model shown in Fig. \ref{CEC-maps}a. These results in
Fig. \ref{QSI-in-plane} and Fig. \ref{LDOS-MS-E} are also in qualitative agreement
with the STM/S\cite{Vershinin04,Hanaguri04,McElroy05,Wise08,Lee09,Schmidt11,Fujita14}
and ARPES\cite{Restrepo23,Shah25,Shen05} experimental observations, where the doping
and energy dependence of the positions of the tips of the Fermi arcs (then the
scattering wave vectors ${\bf q}_{i}$) in the pseudogap phase have been observed. In
particular, the ARPES experiments\cite{Comin16,Comin14,Gh12,Neto14,Campi15,Hash15}
have indicated clearly that in the pseudogap phase, the magnitude of the
charge-order wave vector $Q_{\rm co}={\bf q}_{1}$ smoothly decreases with the
increase of doping, which is also in agreement with the present result shown in
Fig. \ref{LDOS-MS-E}.\\
(iii) As in the case of the SC-state\cite{Zeng24} (see Fig. \ref{QSI-in-plane}f),
the phase $\phi(\bm{q},\omega)$ of the momentum-space LDOS modulation pattern in
the pseudogap phase also exhibits a non-zero phase factor of $\pi$ centered at
around $\bm{q}=\bm{0}$ along the diagonal direction (see Fig. \ref{QSI-in-plane}c),
where the most coherent quasiparticles are located. However, in correspondence with
the spectral weight distribution of the coherent LDOS modulation spectrum in
momentum-space, the phase $\phi(\bm{q},\omega)$ in the SC-state exhibits a far more
coherent behavior than that in the pseudogap phase; \\
(iv) However, in addition to the above QSI octet phenomenology, the exotic
electronic excitation in the pseudogap phase exhibits the LDOS modulation peaks in
momentum-space that are oriented along the parallel direction at the finite
scattering wave vectors ${\bf Q}\approx [\pm 0.36\pi,0]$ and
${\bf Q}\approx [0,\pm 0.36\pi]$. Among the charge-order states observed in the
experiments, the period of the LDOS modulation obtained in the present work at the
charge-order wave vector ${\bf Q}$ corresponds most closely to that the expected
checkerboard charge ordering, which is commonly observed in STM/S\cite{Vershinin04}
and the ARPES\cite{Shen05} measurements on cuprate superconductors in the
pseudogap phase. In the early STM/S\cite{Vershinin04} and ARPES\cite{Shen05}
experiments, the {\it checkerboard charge ordering} was detected in the pseudogap
phase above $T_{\rm c}$. Latter, some STM/S experimental
measurements\cite{Hanaguri04,McElroy05,Wise08,Lee09,Schmidt11,Fujita14} indicated
that although the checkerboard charge ordering appeared in the pseudogap phase
can persist into the SC-state below $T_{\rm c}$, the magnitude of the incoming
photon energy in these experiments is larger than that of the SC gap
($|\omega|>\bar{\Delta}$).
However, whether the checkerboard charge ordering observed in the pseudogap phase or
in the SC-state with the magnitude of the incoming photon energy large compared to
the SC
gap\cite{Vershinin04,Hanaguri04,McElroy05,Wise08,Lee09,Schmidt11,Fujita14,Shen05},
the average values of the checkerboard charge-order wave vectors have been identified
in these experiments as ${\bf Q}\approx [\pm 0.44\pi,0]\pm 15\%$ and
${\bf Q}\approx [0,\pm 0.44\pi]\pm 15\%$, which are qualitatively consistent with
the present results. The checkerboard charge-order state is characterized by the
checkerboard charge-order wave vector ${\bf Q}$, and the qualitative agreement
between theory and experiments in the checkerboard charge-order wave vector
therefore is important to confirm that the checkerboard charge-order state is
generated by the opening of the normal-state pseudogap. More specially, this
checkerboard charge ordering is a static electronic order.
\begin{figure}[h!]
\centering
\includegraphics[scale=0.20]{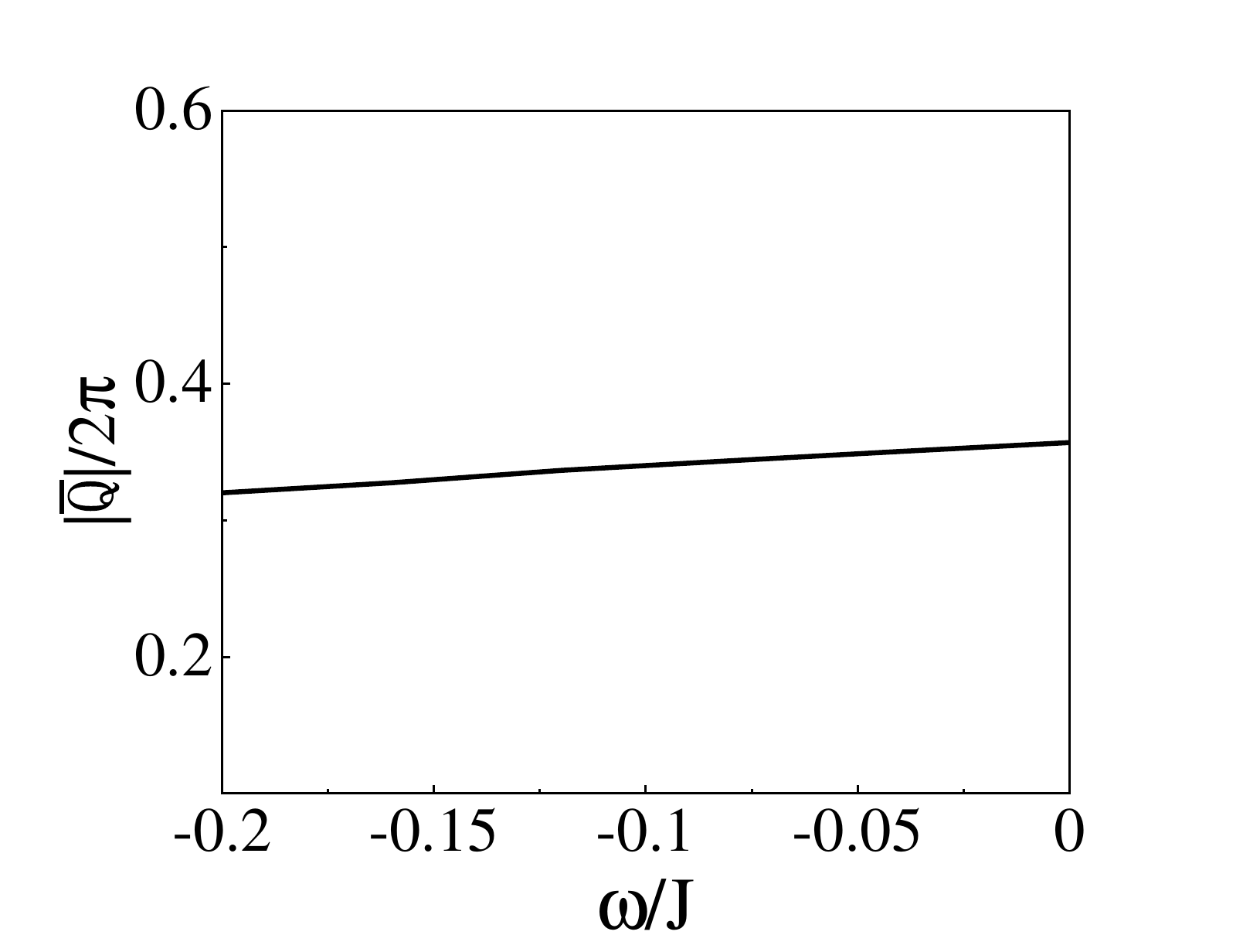}
\caption{The magnitude of the checkerboard charge-order wave vector as a function
of energy for an in-plane single impurity at doping $\delta=0.09$ with temperature
$T=0.07J$ for the impurity-scattering strength $V_{s}=1.0J$ and screening length
$L=5.0a$.\label{Q-energy}}
\end{figure}
To see this unusual
feature more clearly, we have also made a series of calculations for the magnitude
of the checkerboard charge-order wave vector $|{\bf Q}|$ at different energies, and
the result of $|{\bf Q}|$ as a function of energy for an in-plane single impurity at
doping $\delta=0.09$ with temperature $T=0.07J$ for the impurity-scattering strength
$V_{s}=1.0J$ and screening length $L=5.0a$ is plotted in Fig. \ref{Q-energy}, where
$|{\bf Q}|$ is almost independent on energy, confirming that this checkerboard
charge ordering is non-dispersive, also in qualitative agreement with the
experimental observations
\cite{Vershinin04,Hanaguri04,McElroy05,Wise08,Lee09,Schmidt11,Fujita14,Shen05}.

However, in STM/S experiments, the severe systematic error in the direct measurement
of LDOS is produced due to tip elevation errors\cite{Pan01,Fischer07,Yin21}, which
leads to fuzz the LDOS modulation pattern in real-space, and broaden the LDOS
modulation peaks in momentum-space. In this case, it is possible that some intrinsic
aspects of the the quasiparticle excitation in cuprate superconductors may be
masked by the direct measurement of LDOS. Fortunately, with the improvements in the
resolution of STM/S experiments\cite{Kohsaka08,Kohsaka07,Hanaguri07}, the severe
systematic error in the direct measurement of LDOS can be cancelled effectively by
the measurement of the ratio of differential conductances at opposite bias,
\begin{eqnarray}\label{ratio}
Z({\bf r},V)\equiv g({\bf r},+V)/g({\bf r},-V),
\end{eqnarray}
yet retains all the quasiparticle excitation information in the differential
conductance, where $V$ is the bias voltage, and $g({\bf r},V)$ is the differential
conductance. This differential conductance $g({\bf r},\omega)$ is closely related to
the LDOS $\rho({\bf r},\omega)$, and then the intrinsic features of the ratio of
differential conductances at opposite bias can be analyzed theoretically in terms of
LDOS,
\begin{eqnarray}\label{ratio-LDOS}
Z({\bf r},\omega)={\rho({\bf r},\omega)\over\rho({\bf r},-\omega)}.
\end{eqnarray}
Since the condition $\rho_{0}(\omega)\gg\delta\rho({\bf r},\omega)$ is well
satisfied for cuprate superconductors, the ratio of LDOS at opposite energy
(\ref{ratio-LDOS}) can be obtained approximately by the LDOS modulation
$\delta\rho({\bf r},\omega)$ as\cite{ZWang10},
\begin{eqnarray}
Z({\bf r},\omega)\approx Z_{0}(\omega)\left [1+{\delta\rho({\bf r},\omega)
\over\rho_{0}(\omega)}-{\delta\rho({\bf r},-\omega)\over\rho_{0}(-\omega)}\right ],
\end{eqnarray}
where $Z_{0}(\omega)=\rho_{0}(\omega)/\rho_{0}(-\omega)$, and only the first-order
modulation $\delta\rho({\bf r},\pm\omega)$ is kept.
\begin{figure}[h!]
\centering
\includegraphics[scale=0.40]{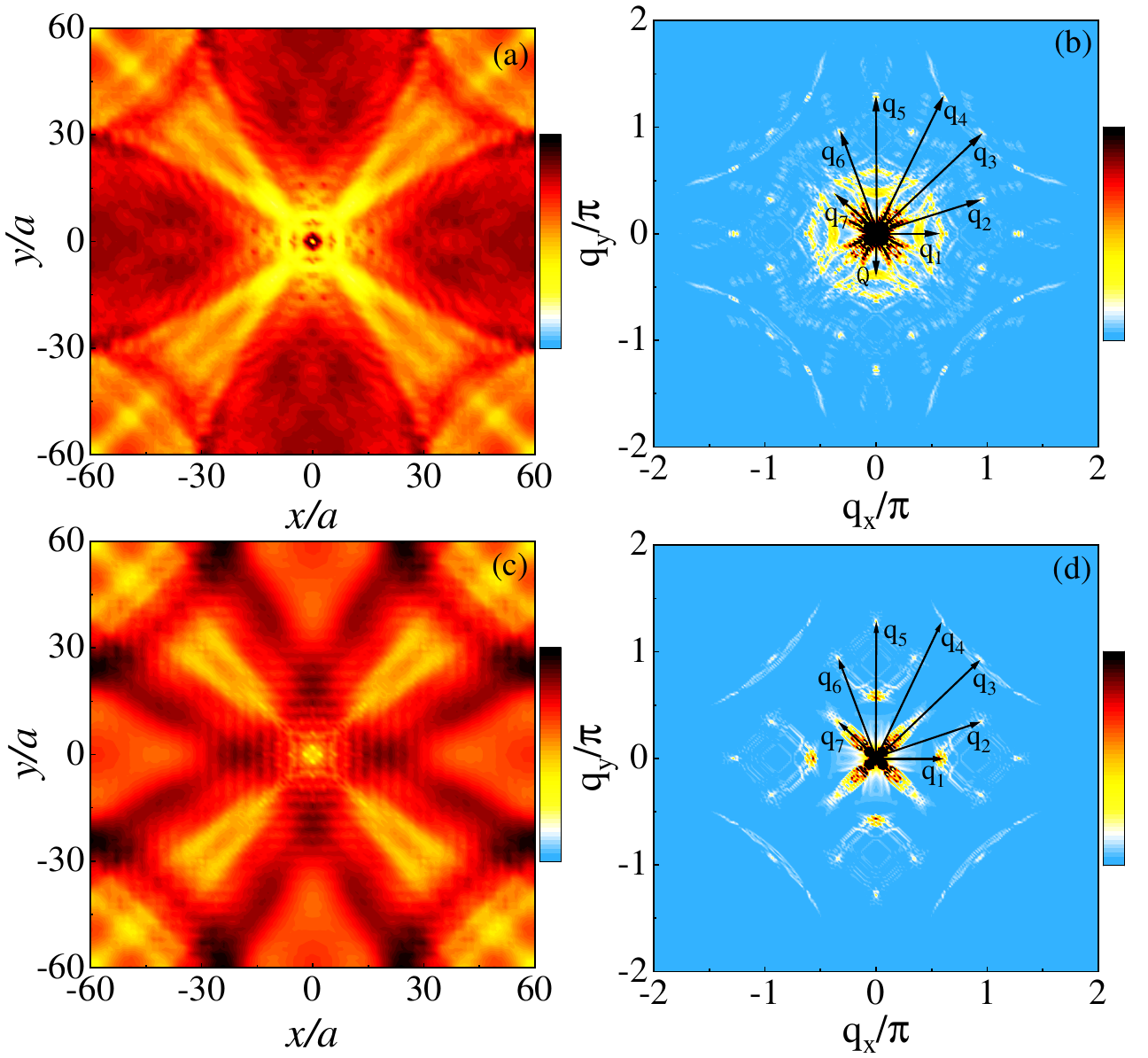}
\caption{(Color online) Upper panel: the intensity maps of the ratio of the local
density of states at opposite energy in the pseudogap phase for an in-plane single
impurity in (a) real-space and (b) momentum-space at doping $\delta=0.09$ with
temperature $T=0.07J$ in the binding-energy $\omega=0.06J$ for the
impurity-scattering strength $V_{s}=1.0J$ and screening length $L=5.0a$. Lower
panel: the corresponding intensity maps of the ratio of the local density of states
at opposite energy in the superconducting-state for an in-plane single impurity in
(c) real-space and (d) momentum-space with temperature $T=0.002J$.
\label{ratio-in-plane}}
\end{figure}
In this case, the modulation of
the ratio of the LDOS at opposite energy can be expressed explicitly as,
\begin{eqnarray}
\delta Z({\bf r},\omega)&=&Z({\bf r},\omega)-Z_0(\omega)\nonumber\\
&\approx& Z_{0}(\omega) \left [{\delta\rho({\bf r},\omega)\over
\rho_{0}(\omega)}-{\delta\rho({\bf r},-\omega)\over\rho_{0}(-\omega)}\right ],
\end{eqnarray}
and its form in momentum-space then is derived straightforwardly by using the
Fourier transformation as,
\begin{eqnarray}\label{equ-z}
\delta Z({\bf q},\omega)\approx Z_{0}(\omega)\left [{\delta\rho({\bf q},\omega)
\over\rho_{0}(\omega)}-{\delta\rho({\bf q},-\omega)\over\rho_{0}(-\omega)}\right ].
\end{eqnarray}
Consequently, this $\delta Z({\bf q},\omega)$ analysis can help to enhance and
unmask minute details of the quasiparticle excitation free from any
set-point-related issues.

To show the unmasked minute details of the quasiparticle excitation more clearly,
we plot the intensity maps of the ratio of the LDOS at opposite energy in the
pseudogap phase in (a) real-space and (b) momentum-space at doping $\delta=0.09$
with temperature $T=0.07J$ in the binding-energy $\omega=0.06J$ for the
impurity-scattering strength $V_{s}=1.0J$ and screening length $L=5.0a$ in the
upper panel of Fig. \ref{ratio-in-plane}, while the corresponding intensity maps
of the ratio of LDOS at opposite energy in the SC-state in (c) real-space and (d)
momentum-space with temperature $T=0.002J$ are plotted in the lower panel of
Fig. \ref{ratio-in-plane}, where the overall features of the $\delta Z$ modulation
are quite similar to the corresponding results of the LDOS modulations shown in
Fig. \ref{QSI-in-plane}, however, there are some substantial differences between
the $\delta Z$ modulation and LDOS modulation: (i) the intensity of the real-space
LDOS modulation along the diagonal direction in Fig. \ref{QSI-in-plane} is
enhanced in the $\delta Z(\bm{r},\omega)$ modulation both in the pseudogap phase
and SC-state, leading to that two ripples observed on the LDOS modulation along
the diagonal direction in Fig. \ref{QSI-in-plane} are converted into a region in
the $\delta Z(\bm{r},\omega)$ modulation. In particular, the vanished weight of
the LDOS modulation connected two ripples along the diagonal direction in the
pseudogap phase shown in Fig. \ref{QSI-in-plane}a is recovered in the
$\delta Z(\bm{r},\omega)$ modulation; (ii) the intensity of the momentum-space
QSI peaks appeared in the LDOS modulation both in the pseudogap phase and
SC-state is enhanced in the $\delta Z(\bm{q},\omega)$ modulation, and then the
QSI peaks becomes more sharper in the $\delta Z(\bm{q},\omega)$ modulation; (iii)
more importantly, in the pseudogap phase, the intensity of the momentum-space
checkerboard peaks emerged in the LDOS modulation in Fig. \ref{QSI-in-plane}b is
strongly enhanced in the $\delta Z(\bm{q},\omega)$ modulation, and then the
fuzziness around checkerboard peaks in the LDOS modulation is eliminated, leading
to that the more sharper and clearer checkerboard peaks appear in the
$\delta Z(\bm{q},\omega)$ modulation. The above results are qualitatively
consistent with the corresponding experimental observations both in the
SC-state\cite{Pan00,Hoffman02,McElroy03,Kohsaka08,Hanaguri09,Vishik09} and the
pseudogap-state\cite{Vershinin04,Hanaguri04,McElroy05,Wise08,Lee09,Schmidt11,Fujita14},
and therefore further confirm that the QSI octet phenomenology both in the pseudogap
phase and SC-state and the checkerboard charge ordering in the pseudogap phase are the
intrinsic phenomena in cuprate superconductors free from any set-point-related issues.

In cuprate superconductors, the impurity distribution accompanied with different types
of the doping processes are quite different\cite{Hussey02,Balatsky06,Alloul09}. In
particular, the atomic dislocation in the charge-carrier doping process gives rise to
impurities in the insulating layers, which often act as smooth scatterers because of
the poor screening effects along the direction normal to the CuO$_{2}$ planes. For
instance, for the cuprate superconductors (Bi,Pb)$_{2}$(Sr,La)$_{2}$CuO$_{6+\delta}$
and Bi$_{2}$Sr$_{1.6}$L$_{0.4}$CuO$_{6+\delta}$ (L=La,Nd,Gd), the mismatch in the
ionic radius between Bi and Pb or Sr and L leads to that the effective impurities
reside in the insulating layers\cite{Eisaki04,Fujita05,Kondo07,Hashimoto08} with
distances away from the conducting CuO$_{2}$ planes, where the concentration of the
out-of-plane impurities is controlled by varying the radius of the Pb or L ions. As in
the case of the SC-state\cite{Zeng24}, the intrinsic phenomena of the QSI octet
phenomenology and the checkerboard charge ordering in the pseudogap phase should be
manifested regardless of the impurity distribution.
\begin{figure}[h!]
\centering
\includegraphics[scale=0.40]{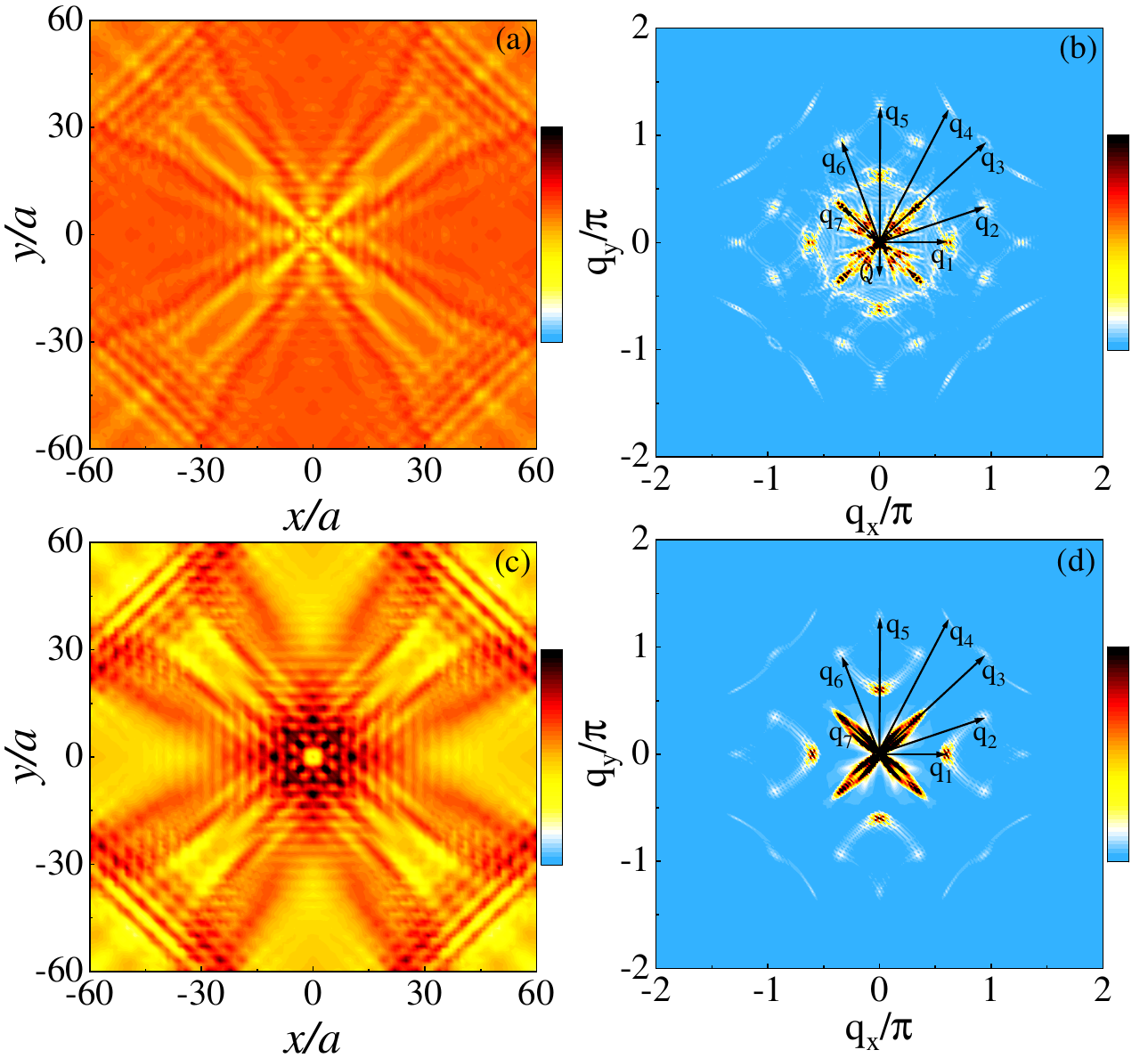}
\caption{(Color online) Upper panel: the intensity maps of the ratio of the local
density of states at opposite energy in the pseudogap phase for an out-of-plane
single impurity in (a) real-space and (b) momentum-space at doping $\delta=0.09$
with temperature $T=0.07J$ in the binding-energy $\omega=0.06J$ for the
impurity-scattering strength $V_{s}=1.0J$ and screening length $L=1.0a$. Lower
panel: the corresponding intensity maps of the ratio of the local density of states
at opposite energy in the superconducting-state for an out-of-plane single impurity
in (c) real-space and (d) momentum-space with temperature $T=0.002J$.
\label{ratio-out-of-plane}}
\end{figure}
To confirm this point
more clearly, we now discuss the modulation of the ratio of LDOS at opposite energy
(then the LDOS modulation) in the presence of an out-of-plane impurity. For an
out-of-plane single impurity located at the lattice site $\bm{r}=0$, the smooth
impurity-scattering potential in real-space can be modeled as\cite{Nunner06},
\begin{equation}\label{VPOT-2-1}
V({\bf r}) = V_{s}e^{-{\bf r}/L},
\end{equation}
with the out-of-plane impurity-scattering strength $V_{s}$ and screening length $L$,
respectively, and then this impurity-scattering potential in momentum-space can be
derived straightforwardly by using the Fourier transformation as,
\begin{equation}\label{VPOT-2}
V_{\bm{k}\bm{k}'}={2\pi V_{s}/L\over [(\bm{k}-\bm{k}')^{2}+1/L^{2}]^{3/2}}.
\end{equation}
In the following discussions, these out-of-plane impurity-scattering strength
$V_{s}$ and screening length $L$ in Eqs. (\ref{VPOT-2-1}) and (\ref{VPOT-2}) are
set as $V_{s}=1.0J$ and $L=1.0a$, respectively.

In the upper panel of Fig. \ref{ratio-out-of-plane}, we plot the intensity maps of
the ratio of the LDOS at opposite energy in the pseudogap phase in (a) real-space
and (b) momentum-space at doping $\delta=0.09$ with temperature $T=0.07J$ in the
binding-energy $\omega=0.06J$ for the impurity-scattering strength $V_{s}=1.0J$ and
screening length $L=1.0a$. For a direct comparison, the corresponding intensity maps
of the ratio of LDOS at opposite energy in the SC-state in (c) real-space and (d)
momentum-space with temperature $T=0.002J$ are plotted in the lower panel of
Fig. \ref{ratio-out-of-plane}. In comparison with the corresponding results obtained
for an in-plane single impurity shown in Fig. \ref{ratio-in-plane}, it thus shows
that the main features of the $\delta Z$ modulations both in the pseudogap phase and
SC-state for an out-of-plane single impurity are essentially the same as the
corresponding ones for an in-plane single impurity. In particular, as a natural
consequence of the quasiparticle scattering geometry that is directly governed by
the shape of the constant energy contour and the related octet scattering model
shown in Fig. \ref{CEC-maps}, (i) the scattering wave vectors $\bm{q}_{i}$ obey the
octet scattering model, in spite of various forms of the impurity scattering
potentials; (ii) the main feature of QSI is similar to that for an in-plane single
impurity as shown in Fig. \ref{ratio-in-plane}; (iii) the essential behaviours of
QSI peak dispersions for an out-of-plane single impurity are almost the same as
those for an in-plane single impurity as shown in Fig. \ref{LDOS-MS-E}, and also
coincide with the evolution of the constant energy contour and the related octet
scattering model in Fig. \ref{CEC-maps} with energy; (iv) the non-dispersive nature
of the checkerboard charge ordering in the pseudogap phase and the related magnitude
of the checkerboard charge-order wave vector $|{\bf Q}|$ are the same as those for
an in-plane single impurity regardless of various forms of the impurity scattering
potentials. However, there are some subtle differences between the $\delta Z$
modulations for an out-of-plane single impurity and an in-plane single impurity:
(i) the real-space $\delta Z(\bm{r},\omega)$ modulations in both the pseudogap
phase and SC-state for an out-of-plane single impurity are more pronounced than the
corresponding ones for an in-plane single impurity, and this difference is also
reflected in the momentum-space $\delta Z(\bm{q},\omega)$ modulations, where the
weights of the QSI peaks both in the the pseudogap phase and SC-state for an
out-of-plane single impurity are larger than these for an in-plane single impurity,
indicating that QSI in the underdoped cuprate superconductors may be generated mainly
by both the strong electron correaltion and out-of-plane single impurity scattering.
On the other hand, the weight of the checkerboard peaks in the pseudogap
phase for an out-of-plane single impurity is smaller than that for an in-plane single
impurity, which shows that the effects both from the strong electron correlation and
the in-plane single impurity scattering play the crucial role for the checkerboard
ordering in the pseudogap phase; (ii) the shape of the $\delta Z(\bm{q},\omega)$
modulation pattern in the pseudogap phase for an out-of-plane single impurity
deviates modestly from the rotundity-like shape of the $\delta Z(\bm{q},\omega)$
modulation pattern for an in-plane single impurity. These results therefore confirm
that the essential features of the QSI octet phenomenology both in the pseudogap
phase and SC-state as well as the checkerboard charge ordering in the pseudogap phase
are the intrinsic phenomena regardless of the impurity distribution.

In the recent discussion of the nature of QSI in the SC-state\cite{Zeng24}, we
have shown that the overall feature of the LDOS modulation in the SC-state can be
depicted qualitatively by taking into account the quasiparticle scattering from a
single impurity on the kinetic-energy-driven homogeneous SC-state, however, the
pronounced QSI peaks in the momentum-space LDOS modulation for a single impurity are
smeared heavily for multiple impurities, and then the momentum-space LDOS modulation
for multiple impurities exhibits a speckle pattern. In this case, we have also
discussed the LDOS modulation in the pseudogap phase for multiple impurities, and the
obtained results show that as in the case of the SC-state, although the pronounced
QSI peaks and the checkerboard peaks in the momentum-space LDOS modulation for a
single impurity are also smeared heavily for multiple impurities, the characteristic
structure of the LDOS modulation spectrum are essentially the same both for multiple
impurities and a single impurity.

Now we try to give an explanation to show (i) why the characteristic of QSI observed
in the SC-state remains unchanged upon passing above $T_{\rm c}$ in the pseudogap
phase? and (ii) why the non-dispersive checkerboard charge ordering emerges in the
pseudogap phase only? The nature of the LDOS modulation is mainly governed by the
momentum-space profiles of the joint density of states. Although the SC-state of the
underdoped cuprate superconductors coexists and competes with the pseudogap-state,
the SC-state quasiparticle still exhibits the particle-hole
mixing\cite{Campuzano96,Matsui03} similar to that in the conventional
superconductors\cite{Bardeen57,Schrieffer64}. In particular, in the low-energy limit,
the quasiparticle energy dispersion $\varpi_{\bf k}(\omega)$ in Eq. (\ref{QPED}) in
the SC-state can be reduced as\cite{Li25}, $\varpi_{\bf k}=E_{\bf k}$, where
$E_{\bf k}=\sqrt{\bar{\varepsilon}^{2}_{\bf k}+\mid\bar{\Delta}_{{\rm Z}{\bf k}}
\mid^{2}}$, with $\bar{\varepsilon}_{\bf k}=Z_{\rm F}\varepsilon_{\bf k}$, and
$\bar{\Delta}_{{\rm Z}{\bf k}}=Z_{\rm F}\bar{\Delta}_{\bf k}$, while the
single-particle coherent weight $Z_{\rm F}$ that is directly associated with the
normal-state pseudogap has been given explicitly in Ref. \onlinecite{Li25}. In this
case, the scattering wave vector ${\bf q}_{i}$ obeying the octet scattering model
dominates the LDOS modulation at energy $\omega$, since these ${\bf k}$-pairs on
the constant energy contour shown in Fig. \ref{CEC-maps}b connected by the
scattering wave vector ${\bf q}_{i}$ have a high joint density of states. This also
follows from a basic fact that the magnitude of the SC gap
$\mid\bar{\Delta}_{\bf k}\mid$ is momentum-dependent, and then some different minima
$\omega_{i}$, at which the quasiparticles emerge, occur at the tips of the Fermi
arcs\cite{Pan00,Hoffman02,McElroy03,Kohsaka08,Hanaguri09,Vishik09}. However, the
square-lattice CuO$_{2}$ plane has a four-fold rotation symmetry, which implies that
at any non-zero energy $\omega_{i}$, there are only eight possible ${\bf k}_{\rm F}$
values at which $\mid\bar{\Delta}_{{\rm Z}{\bf k}}\mid=\omega_{i}$ in the first
Brillouin zone\cite{Pan00,Hoffman02,McElroy03,Kohsaka08,Hanaguri09,Vishik09}. The
highest joint density of states for the quasiparticle scattering at this energy
$\omega_{i}$ occur at ${\bf q}_{i}$-vectors connecting the eight tips of the Fermi
arcs. This is why the SC-state QSI peaks in the momentum-space LDOS modulation occur
at these ${\bf q}_{i}$-vectors. However, the existence of the QSI octet phenomenology
in the pseudogap phase is not very surprised, since all the behaviours of the QSI
octet phenomenology observed in the STM/S experiments are closely related to the
momentum-space profiles of the joint density of
states\cite{Chatterjee06,McElroy06,Chatterjee07,He14,Restrepo23,Shah25}, while the
main feature of the momentum-space profiles of the joint density of states in the
pseudogap phase\cite{Vershinin04,Hanaguri04,McElroy05,Wise08,Lee09,Schmidt11,Fujita14}
characterized by the octet scattering model shown in Fig. \ref{CEC-maps}a is almost
the same as that in the
SC-state\cite{Pan00,Hoffman02,McElroy03,Kohsaka08,Hanaguri09,Vishik09} shown in
Fig. \ref{CEC-maps}b. In other words, the main feature of the joint density of states
occurred in the SC-state survive virtually unchanged into the pseudogap phase, which
leads to that the SC-state QSI octet phenomenology exists in the pseudogap phase.
However, the spectral weight at around the antinodal region in the SC-state is gapped
out by both the normal-state pseudogap and the d-wave SC gap, while the spectral
weight at around the antinodal region in the pseudogap phase is suppressed partially
by the normal-state pseudogap only, resulting in the existence of small residual
weight at around the antinodal region in the pseudogap phase. In this case, the
checkerboard charge ordering in the SC-state is suppressed strongly by both the
normal-state pseudogap and SC gap, leading to that the checkerboard charge ordering
in the SC-state becomes unobservable in experiments, while this checkerboard charge
ordering in the pseudogap phase is suppressed partially by the normal-state pseudogap,
leading to that the checkerboard charge ordering in the pseudogap phase becomes
observable in experiments. This is why the checkerboard peaks in the momentum-space
LDOS modulation occur in the pseudogap phase only.

\section{Summary}\label{Summary}

Within the framework of the kinetic-energy-driven superconductivity, we have
rederived the homogeneous electron propagator, where the strong coupling between
electrons directly from the kinetic energy by the exchange of the spin excitation
induces the pseudogap-state in the particle-hole channel and the SC-state in the
particle-particle channel. The pseudogap and the SC gap respectively originate
from the electron normal self-energy in the particle-hole channel and the
electron anomalous self-energy in the particle-particle channel, and are obtained
explicitly by taking into account the vertex correction. This pseudogap-state
coexists and competes with the SC-state below $T_{\rm c}$, leading to a dome-like
shape of the doping dependence of $T_{\rm c}$, and the exotic pseudogap phase
with the normal-state pseudogap persisting up to $T^{*}$ that is substantially
higher than $T_{\rm c}$ in the underdoped regime. Moreover, the structure of the
constant energy contour intensity maps in the pseudogap phase essentially
coincides with the corresponding one in the SC-state, where the pseudogap
suppresses strongly the spectral weight on the constant energy contour at around
the antinodal region, while it suppresses modestly the spectral weight at around
the nodal region, leading to the formation of the Fermi arcs centered around the
nodal region. These tips of the Fermi arcs connected by the quasiparticle
scattering wave vectors ${\bf q}_{i}$ naturally construct an octet scattering
model. Starting from the microscopic homogeneous electron propagator and the
related octet scattering model, we have studied the nature of the LDOS modulation
in the pseudogap phase of cuprate superconductors within the $T$-matrix approach,
where the $T$-matrix generated by the quasiparticle scattering from various kinds
of a single impurity is derived accurately, and is employed to derive LDOS by the
involvement of all the quasiparticle excitations and scattering processes. Our
results show a number of anomalous electronic-state properties in the pseudogap
phase of cuprate superconductors is directly correlated to the
opening of the normal-state pseudogap: (i) as a natural consequence of the same
characteristic of the octet scattering model both in the SC-state and pseudogap
phase, the QSI octet phenomenology in the SC-state survive virtually unchanged
into the pseudogap phase; (ii) however, the electronic density of states on the
constant energy contour at around the antinodal region in the SC-state is gapped
out completely by both the d-wave SC gap and normal-state pseudogap, while it in
the pseudogap phase is depressed partially by the normal-state pseudogap, which
directly leads to that the non-dispersive checkerboard charge ordering with a
finite wave vector ${\bf Q}$ emerges in the pseudogap phase only. Our results in
this paper together with the previous work for QSI in the SC-state\cite{Zeng24}
therefore show that the electronic-state affected by the normal-state pseudogap
exhibit the LDOS modulation spectrum organization.

\section*{Acknowledgements}

X.L. and S.P. are supported by the National Key Research and Development Program of
China under Grant Nos. 2023YFA1406500 and 2021YFA1401803, and the National Natural
Science Foundation of China (NSFC) under Grant No. 12274036. M.Z. is supported by
NSFC under Grant No. 12504172 and the Special Funding for Postdoctoral Research
Projects in Chongqing under Grant No. 2024CQBSHTB3156. H.G. acknowledge support from
NSFC under grant No. 12574249.

%\newpage

\appendix

\begin{widetext}

\section{Derivation of $T$-matrix in the pseudogap phase}
\label{Derivation-of-T-matrix}

In this Appendix, we follow the previous derivation of the $T$-matrix in the
SC-state\cite{Zeng24} to evaluate accurately the $T$-matrix in the pseudogap phase.
The $T$-matrix in Eq. (\ref{norman-T-matrix-equation-1}) of the main text can be
expressed explicitly in terms of the equation of iteration method as,
\begin{eqnarray}\label{norman-T-mat-eq}
T_{\bm{k}\bm{k}'}(\omega)={1\over N}V_{\bm{k}\bm{k}'}+{1\over N}\sum_{\bm{k}_{1}}
V_{\bm{k}\bm{k}_{1}}G(\bm{k}_{1},\omega){1\over N}V_{\bm{k}_{1}\bm{k}'}+{1\over N}
\sum_{\bm{k}_{1}}V_{\bm{k}\bm{k}_{1}}G(\bm{k}_{1},\omega){1\over N}\sum_{\bm{k}_{2}}
V_{\bm{k}_{1}\bm{k}_{2}}G(\bm{k}_{2},\omega){1\over N}V_{\bm{k}_{2}\bm{k}'}+\cdots,
~~~~~
\end{eqnarray}
which can be derived straightforwardly as\cite{Zeng24},
\begin{equation}\label{Tmat-Solution}
T_{\bm{k}\bm{k'}}(\omega)=\left ( \bar{V}{1\over 1-\bar{M}}\right )_{\bm{k}\bm{k'}},
\end{equation}
where $\bar{V}=V/N$ is a matrix, $N$ is the number of lattice sites, while the matrix
$\bar{M}$ can be obtained explicitly as,
%\begin{widetext}
\begin{eqnarray}
\bar{M}&=&
\left(\begin{array}{ccccc}
G(\bm{k}_{1},\omega)\bar{V}_{\bm{k}_{1}\bm{k}_{1}} &
G(\bm{k}_{1},\omega)\bar{V}_{\bm{k}_{1}\bm{k}_{2}} &
G(\bm{k}_{1},\omega)\bar{V}_{\bm{k}_{1}\bm{k}_{3}} & \cdots &
G(\bm{k}_{1},\omega)\bar{V}_{\bm{k}_{1}\bm{k}_{N}} \\
G(\bm{k}_{2},\omega)\bar{V}_{\bm{k}_{2}\bm{k}_{1}} &
G(\bm{k}_{2},\omega)\bar{V}_{\bm{k}_{2}\bm{k}_{2}} &
G(\bm{k}_{2},\omega)\bar{V}_{\bm{k}_{2}\bm{k}_{3}} & \cdots &
G(\bm{k}_{2},\omega)\bar{V}_{\bm{k}_{2}\bm{k}_{N}} \\
\vdots & \vdots & \vdots & \vdots & \vdots\\
G(\bm{k}_{N},\omega)\bar{V}_{\bm{k}_{N}\bm{k}_{1}} &
G(\bm{k}_{N},\omega)\bar{V}_{\bm{k}_{N}\bm{k}_{2}} &
G(\bm{k}_{N},\omega)\bar{V}_{\bm{k}_{N}\bm{k}_{3}} & \cdots &
G(\bm{k}_{N},\omega)\bar{V}_{\bm{k}_{N}\bm{k}_{N}}
\end{array}\right).
\end{eqnarray}
%\end{widetext}
However, the $T$-matrix in Eq. (\ref{norman-T-matrix-equation-1}) of the main text
can be also derived directly as,
\begin{equation}
T(\omega) = \bar{V} + \bar{M}(\omega)T(\omega) = \frac{1}{1-\bar{M}(\omega)}\bar{V},
\end{equation}
with the matrix,
%\begin{widetext}
\begin{eqnarray}
\bar{M}&=&
\left(\begin{array}{ccccc}
\bar{V}_{\bm{k}_{1}\bm{k}_{1}}G(\bm{k}_{1},\omega) &
\bar{V}_{\bm{k}_{1}\bm{k}_{2}}G(\bm{k}_{2},\omega) &
\bar{V}_{\bm{k}_{1}\bm{k}_{3}}G(\bm{k}_{3},\omega) & \cdots &
\bar{V}_{\bm{k}_{1}\bm{k}_{N}}G(\bm{k}_{N},\omega) \\
\bar{V}_{\bm{k}_{2}\bm{k}_{1}}G(\bm{k}_{1},\omega) &
\bar{V}_{\bm{k}_{2}\bm{k}_{2}}G(\bm{k}_{2},\omega) &
\bar{V}_{\bm{k}_{2}\bm{k}_{3}}G(\bm{k}_{3},\omega) & \cdots &
\bar{V}_{\bm{k}_{2}\bm{k}_{N}}G(\bm{k}_{N},\omega) \\
\vdots & \vdots & \vdots & \vdots & \vdots\\
\bar{V}_{\bm{k}_{N}\bm{k}_{1}}G(\bm{k}_{1},\omega) &
\bar{V}_{\bm{k}_{N}\bm{k}_{2}}G(\bm{k}_{2},\omega) &
\bar{V}_{\bm{k}_{N}\bm{k}_{3}}G(\bm{k}_{3},\omega) & \cdots &
\bar{V}_{\bm{k}_{N}\bm{k}_{N}}G(\bm{k}_{N},\omega)
\end{array}\right).
\end{eqnarray}
\end{widetext}

\end{document}